\RequirePackage{fix-cm}
\documentclass{svjour3}                     % onecolumn (standard format)
\smartqed  % flush right qed marks, e.g. at end of proof
\usepackage{subfigure}
\usepackage{booktabs}
\usepackage{graphicx}
\usepackage{natbib}
\usepackage{lineno}
\usepackage{latexsym}
\usepackage{color}

\newcommand\red[1]{{\color{red}#1}}

 \journalname{Climate Dynamics}
\begin{document}

\title{Analysis of  rainfall  seasonality from observations and climate models  %\thanks{Grants or other notes
}


\author{Salvatore Pascale         \and
        Valerio Lucarini    \and
         Xue Feng    \and
          Amilcare Porporato  \and
           Shabeh ul Hasson
}


\institute{S. Pascale, V. Lucarini, S. Hasson \at
              Meteorologisches Institut, Center for Earth System Research and\\ Sustainability (CEN),  Universit\"at Hamburg, Grindelberg 5, \\Hamburg, 20144, Germany.\\
              Tel.: +49 (0)40 42838 9207\\
              \email{salvatore.pascale@uni-hamburg.de}           %  \\
           \and
           V. Lucarini \at
            Department of Mathematics and Statistics \\ University of Reading, Reading, UK
           \and
           X.Feng, A. Porporato \at
           Department of Civil and Environmental Engineering \\ Duke University, North Carolina, USA   
}

\date{Received: date / Accepted: date}

\maketitle

\begin{abstract}

Two new  indicators of rainfall seasonality based on information entropy, the  relative entropy (RE) and the  dimensionless seasonality index (DSI), together with  the mean annual rainfall,  are evaluated  on a global scale for  recently updated precipitation gridded datasets   and for historical simulations from coupled atmosphere-ocean general circulation models. The RE provides a measure of the number of wet months and,  for precipitation regimes featuring one maximum in the monthly rain distribution,  it is related  to the duration of the wet season. The DSI combines  the rainfall intensity  with its degree of  seasonality and it is an indicator of the extent  of the  global monsoon region.  We show that  the RE and the DSI are fairly  independent of the time resolution of the precipitation data,  thereby allowing   objective metrics for model intercomparison and ranking. 
 Regions with different precipitation regimes are classified and characterized in terms of RE and DSI.   Comparison of different land observational  datasets reveals  substantial difference  in their local  representation of seasonality.  It is shown that   two-dimensional  maps of RE provide an easy way to compare rainfall seasonality from  various datasets and to determine areas of interest.  CMIP5  models consistently  overestimate the RE over  tropical Latin America and  underestimate it in  Western Africa and  East Asia.  It is demonstrated that      positive   RE biases in a GCM   are associated with  simulated  monthly precipitation fractions which are   too large during the wet months and too small in the months preceding the wet season;  negative biases are instead  due to an excess of rainfall during the dry months.

\keywords{Rainfall seasonality \and information entropy \and  hydrological cycle \and CMIP5 models}
\end{abstract}

\section{Introduction}
\label{intro}

The  increase of greenhouse gases in the atmosphere is substantially altering the Earth's energy budget  and  warming   the climate system  \citep{IPCC5}. One of the most crucial aspect that we need to understand and quantify   is how     greenhouse gas forcing   is going to impact, globally and locally,  the hydrological cycle and the precipitation patterns over the globe. Precipitation, in particular, plays a key role in the hydrological cycle and it is one of the climate variables of the highest concern among climate scientists. 

In spite of the difficulty of monitoring a field highly variable in both  space and time  such as precipitation, changes of the annual mean precipitation have    been detected in observations and attributed to human influences   \citep{Zhang, Noake, Beena}.  Simulations performed by  coupled general circulation models (GCMs)    predict  that  the large-scale  hydrological cycle is  affected by climate warming in a complex  way.  Thermodynamical effects \citep{Soden, Allen, Meehl,  Chou2} -- associated with an  increase of specific humidity   -- and dynamical effects -- related to  changes of the large scale tropical circulation and moisture transport due to baroclinic eddies  and tropical circulation \citep{Seager, Camargo} --  both contribute to the   changing global  patterns of precipitation \citep{ Chadwick}. The first approximation  emerging pattern of changes in the hydrological cycle   indicates that subtropical arid and semi-arid regions  are expected to get drier \citep[e.g.][]{ Kelley, Seager2}  whereas wet equatorial  and high-latitude regions are expected  to get wetter.  Understanding  future precipitation changes both in the tropics and extratropics  is a challenge for climate science because   it   requires  knowing      how  different  large-scale weather systems such as the  monsoons \citep{Vecchi, Cherchi, Turner,  Sperber, Kitoh, Cook, Shabeh1, Shabeh2}, the Hadley Cell \citep{Kang},  midlatitude  baroclinic cyclones \citep{Bengtsson, Harvey, Zappa} and tropical cyclones \citep{Rathmann} will change under greenhouse gas forcing.  While there is modeling evidence that storm tracks shifts polewards as the climate warms globaly \citep{Bengtsson, Swart},  GCMs still faces  serious difficulty in simulating   the regional distribution of monsoons rainfall under present conditions and tend to disagree on future projections \citep{Turner}.  These uncertainties make the use of GCMs projections problematic  for applications, such as the  assessment of rivers hydrology \citep{Danube, Shabeh1}.
  
For a more  robust GCMs validation and for a  more complete description  of the precipitation regimes  under global warming, it is  important   to take into account not only   the mean  total annual amount of  precipitation but also   statical properties of intense  rainfall events \citep{Sillman, Kharin, Mehran} and  
  the seasonality of the annual rainfall \citep{Porporato}. The latter  will be the topic of this study. A complete  description of  rainfall seasonality needs to quantify    the  duration of the wet and dry seasons, their intensity and their timing \citep{Chou, Noake,Sperber, Shabeh2}.   Particularly in the tropics, ecosystems are extremely sensitive to the arrival of rain at the beginning of the wet season and to  the wet season length \citep{Borchert, Eamus, Rohr, Itur}.   Furthermore,  precipitation  seasonality, with its related drought and flood risks, makes agricultural efforts and sustainable management  of water resources more problematic, posing a challenge for local populations.   It is therefore of key importance to quantify how the seasonality of precipitation is changing in a warming climate.

Traditionally,  rainfall seasonality is   investigated through  latitude-months  Hovm\"oller diagrams \citep[e.g.][]{Seth, Sperber, Huang}, showing  how  zonally averaged rainfall   evolves during the year at  each latitude within a certain study area. Although this is certainly a natural  way to study rainfall seasonality, such an approach cannot  provide  detailed information  (e.g. two-dimensional map    of rainfall seasonality)      and  cannot be used, for example, to study interannual variability and long-term timeseries, which instead requires a local (i.e. dependent on latitude and longitude) scalar measure of seasonality.       Several loosely related but not equivalent metrics  -- such as the relative lengths and rainfall amounts of the wet and dry seasons and the arrival dates of the 25th and 75th percentile rainfall \citep[e.g.][]{Walsh} -- can be found in the literature, often lacking general applicability \citep{Shukla, Webster, Wang3, Goswami, Kajikawa}.    A  meaningful comparison of the rainfall seasonality of different locations   and of different periods (e.g. 21st century projections) or between different models  requires a precise and robust  quantification of this aspect of rainfall regimes.

 Novel   seasonality indicators of precipitation regimes -- the  relative entropy (RE) and the dimensionless seasonality index (DSI) -- have been recently introduced by \cite{Porporato} based on    the definition of relative entropy \citep[e.g.][]{Cover}  and applied to tropical regions between $20^\circ $ N/S for detecting changes in rainfall seasonality in the tropics.    While  relative entropy  is well known  in statistical  physics and information theory,  its use made  for precipitation seasonality analysis is new.   Such an approach relies on quantifying the differences between  the time series, for a given year,    of the monthly fraction of the annual precipitation $p_m=r_m/R$ ($r_m$ is the precipitation accumulated in the $m$th month and $R=\sum_{m=1}^{12} r_m$ the total annual precipitation) and  the   uniform precipitation sequence $q_m=1/12$. 
 %By adding information on the average yearly precipitation,  a relatively  complete picture of the properties of the monthly rainfall of a region is obtained. 
 Such an approach is very general and does not rely on specific assumptions or on arbitrary thresholds, thus provides   new, well-founded metrics  \citep[e.g.][]{Knutti} for evaluating the capability of climate models in simulating rainfall regimes.
 
In this study we       estimate, for the first time, the seasonality indicators introduced by \cite{Porporato} also over oceans and outside the Tropics   and  compare      observations   and GCMs simulations over the historical period 1950-2010. The goal is to show how these  newly introduced metrics can be used     for a systematic characterization of seasonality of global precipitation regimes.  In particular in this paper we will:
\begin{itemize}
\item[1)] characterize  precipitation regimes in terms of the new indexes;
\item[2)] assess the capability of CMIP5   coupled ocean-atmosphere  models in reproducing them; 
\item[3)] show how these indexes can be  used to detect models' deficiencies in simulating  the seasonal cycle of precipitation. 
\end{itemize}
The use of   the RE will allow us   to compare very easily  seasonality of observational and simulated precipitation datasets (CMIP5) and to detect 
models deficiencies  in representing  the rainfall seasonal cycle.
The paper is structured in the following way. In Sect.~\ref{data_methods} the datasets used for our analysis and the methods for estimating indexes are explained. In Sect.~\ref{clim} the climatology of such indicators is  presented and discussed in the context of atmospheric general circulation.  The capability of coupled global models to simulate the RE    is assessed  in Sect.~\ref{intercomp}    and the main findings summarized in Sect.~\ref{finale}. In Appendix   a  summary of the main properties of the various indexes based on relative  entropy    is given.

\section{Data and Methods}
\label{data_methods}

\subsection{Observational data}
\label{dataset}

 Two newly updated land precipitation  datasets  are used  in this study: a) the updated gridded climate dataset developed at the Climatic Research Unit,  CRU TS3.10, simply refereed to as CRU in the following  \citep{CRU, Mitchell} and b) the Global Precipitation Climatology Centre dataset, referred to as GPCC \citep{GPCC, GPCC2}. GPCC and CRU reanalysis are based on statistically interpolated  \emph{in situ} rain measurements   and cover all land areas -- except Antartica -- at   monthly temporal resolution  for the period $1901-2010$. GPCC precipitation fields  are available  on grids of  different angular resolutions ($0.5 ^\circ \times 0.5 ^\circ $, $1^\circ \times 1^\circ$ and $2.5^\circ \times 2.5^\circ$) whereas the CRU dataset is available  at $0.5^\circ$. 
In addition, to estimate  the indicators  climatology  over the whole globe, including the oceans,   the Climate Prediction Center Merged Analysis of Precipitation dataset \citep[CMAP, ][]{Xie} and the Global Precipitation Climatology Project monthly precipitation dataset  \citep[GPCP,][]{Xie2},   available at monthly temporal resolution (1979-2009) and at   $2.5^\circ \times 2.5^\circ$ degrees,   will also be used in this study.
CMAP and GPCP  datasets are compiled from merged satellite precipitation data  and bias-corrected over land through continental rain-gauge observations \citep{Bolvin, Huffman}.    GPCC, CRU,  CMAP and GPCP  have been validated  and used   in numerous studies focusing on the hydrological cycle including   both global \citep[e.g.][]{Kitoh, Chou, Frierson} and     regional analysis \citep[e.g.][]{Cook, Sperber}.  It is worth  mentioning  here that, because of  the uneven spatial and temporal coverage of the  gauging stations, terrain heterogeneities  and uncertainties added by quality control and interpolation techniques,  combined with the complex spatial and temporal variability of precipitation at all scales \citep{Lovejoy, Deidda},  caution is needed when interpreting the results based on such gridded datasets, including upscaled quantities. Because of these reasons,  reconstructed precipitation datasets are much more uncertain than   reconstruction of smoother and more regular fields such as, for example, surface temperature.

\subsection{Model simulations}
\label{models}
We analyze  the climate models data produced for the IPCC 5th Assessment Report  \citep{IPCC5} and collected through the  Coupled Model Intercomparison Project platform,  Phase 5 \citep[CMIP5, ][]{Taylor, Gui} for  the monthly  means of  precipitations.      Table~\ref{tab1} shows the basic information about  the CMIP5 models considered  in this study such as horizontal and vertical spatial resolution of the atmospheric  models and the research institutions where models have been developed.   The  CMIP5 database contains   long-terms runs for simulation of the industrial period (from mid-ninenteeth century to present) and future climate projections according to different emission  scenarios  (``representative concentration pathways"  RCPs, \cite{RCP}). Simulations of the historical period 1850-2005  forced with both anthropogenic and natural forcings  are used in this study.   Most of the CMIP5 models selected here  provide multiple ensemble  members for each considered scenario. Here we  selected just the first member of the ensemble for each model.     GCMs with serious inconsistencies in the water cycle --  i.e.  models in which     long-term annual means of evaporation minus precipitation is   larger, in absolute value,  than  $10^5$\,m$^3$\,s$^{-1}$  and equivalent to a latent energy  bias larger than  $\approx 1$ W\, m$^{-2}$  \citep{Liepert2} -- have been left out from our analysis.
 
In order to make a  spatial  intercomparison between  models and observations, models precipitation, RE and DSI fields are estimated on each models' own horizontal grid,   then   multiplied by their own land sea mask and finally  linearly interpolated on a $1^\circ \times 1^\circ$ horizontal grid, to compare with GPCC and CRU,  or on a $2.5^\circ \times 2.5^\circ$ for comparison with the CMAP or GPCP dataset. Because of the sparseness of observed  data before $1950$ \citep{GPCC},   our climatological and trend analysis will be restricted to the $1950-2010$ period for both observations and models. Results on  seasonality changes  in  future climate scenarios warming climates for the next century  under different representative concentration pathways will be presented elsewhere.

%For the model selection I can also have a look at other studies (e.g that of \cite{Kumar})

\subsection{Relative entropy and dimensionless seasonality index}
\label{methods}

%\paragraph{Seasonality indicators}
Given a  monthly precipitation frequency   at the surface point $x=(\phi, \lambda)$ ($\phi$ latitude, $\lambda$  longitude) associated with   monthly precipitations  $r_m(x)$, $m=1,\ldots 12$ 
\begin{equation}
p_m(x)=r_m(x)/R(x), \quad R(x)=\sum_{m=1}^{12} r_m(x),
\label{prob}
\end{equation}
the relative entropy  
\begin{equation}
D(p(x)) = \sum_{m=1}^{12} p_m(x)\log_2\left( 12\,p_m(x) \right)
\label{again}
\end{equation}
 is a measure of the inefficiency of assuming that the distribution is  the monthly uniform precipitation sequence  $q_m=1/12$, $m=1,\ldots,12$ when the true distribution is $p$ and it is a way to   quantify how different $p$ is  from $q$. The relative entropy    is closely  related to the number of ``wet'' months (see Appendix) and it reaches   its maximum value $\log_2 12$ when the annual rainfall is concentrated in one single month (the limit $p_m\log_2 p_m\rightarrow 0$   is taken for $p_m=0$) and  equal to zero for  $p=q$.  
 
Given $D(x)$, the dimensionless seasonality index (DSI) is  defined as \citep{Porporato}
\begin{equation}
S(x)=D(x)\left(\frac{R(x)}{R_0}\right)
\label{dsi}
\end{equation}
in which $R_0$ is a constant  scaling factor introduced in order to make the precipitation dimensionless. We choose $R_0$ as the maximum of $R(x)$ of  gridded datasets over the whole period $1950-2010$ and equal to $9701$ mm.  According to definition (\ref{dsi}), $S$ is zero when either $R$ (completely dry location) or $D$ ($R$ distributed uniformly throughout the year) are zero and maximum ($\log_2 12$) when $R_0$ is concentrated in a single month.   For our analysis we compute $R_k(x)$,  $D_k(x)$ and $S_k(x)$  for each hydrological   year $k$ from the rainfall probability  $p_{m,k}(x)=r_{m,k}/R_k$  ($k$th year,   $m$th month)  and then take the climatological mean $\overline{(\cdot)}$ over a certain time period, that is $\overline{R}(x)=\overline{R_k(x)}$, $\overline{D}(x)=\overline{D_k(x)}$, $\overline{S}(x)=\overline{S_k(x)}$.  
A succinct   summary of the main properties of the various indexes based on relative  entropy    is given in Appendix.

%Results are insensitive to whether the beginning of each year is in January  (which is the natural definition for regional monsoons in the northern hemisphere) or July (natural definition for  monsoons  in southern hemisphere).

\section{Analysis of observed  climatology}
\label{clim}

\subsection{Global patterns of RE and DSI}

\begin{figure*}
 \centering
 \subfigure{\includegraphics[angle=0, width=1.2\textwidth]{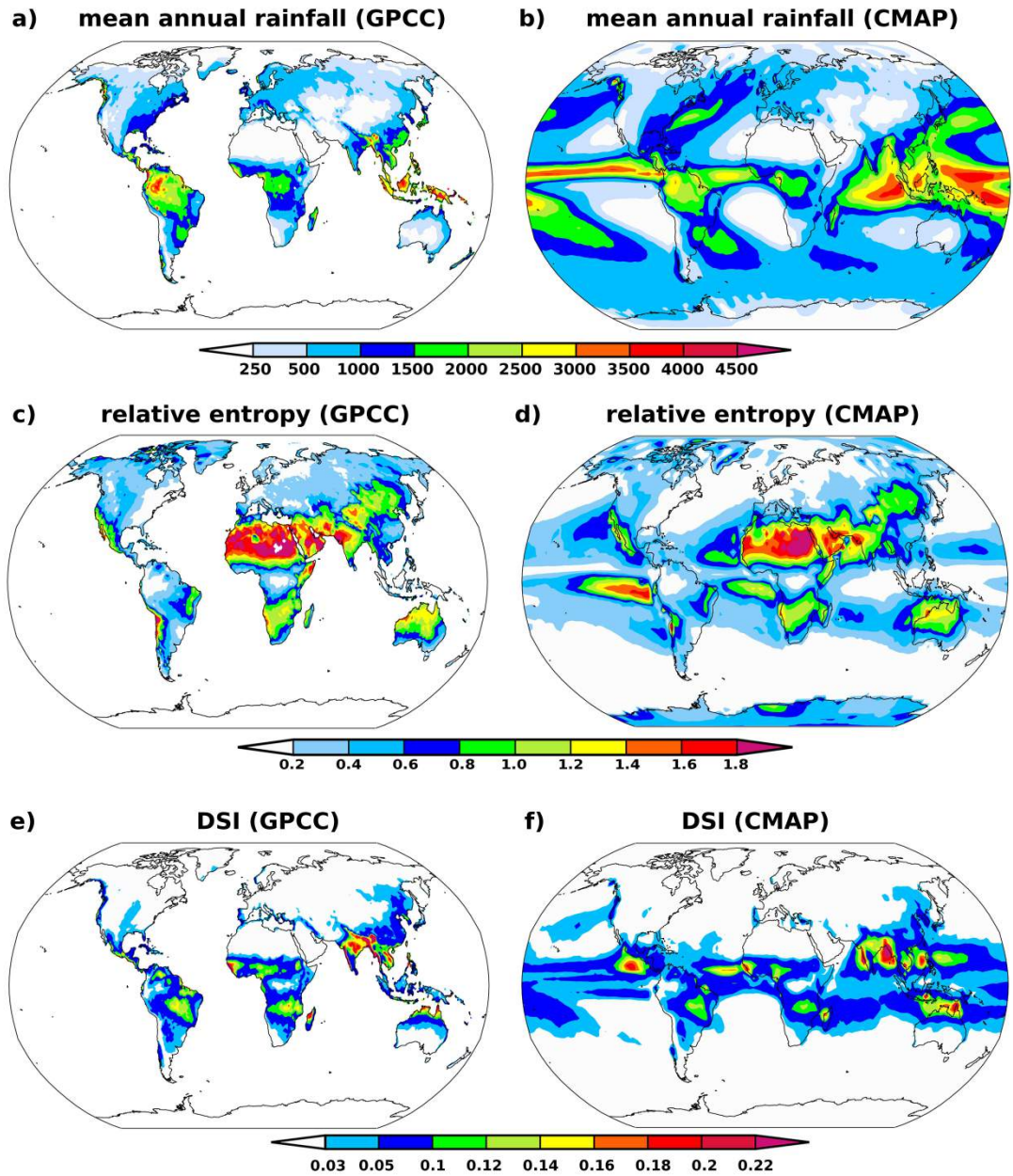} \label{fig1} }
\caption{(a, b) Mean annual rainfall  $\overline{R}$ (mm\,yr$^{-1}$),   (c, d) relative entropy  $\overline{D}$  and  (e, f) dimensionless seasonality index $\overline{S}$    for the GPCC land dataset (left column) and CMAP dataset  (right column).   \label{seconda} }
\end{figure*}

In Fig.~\ref{seconda} mean annual precipitation,  relative entropy and the dimensionless seasonality index  for the GPCC land dataset are shown over the period 1950-2010.   Values  over oceans are shown on the right side of   Fig.~\ref{seconda}  by using the updated 1979-2009 CMAP dataset  \citep{Xie}. Regions with  the largest  $D$  are those placed in the subtropical zone between $10^\circ$ and $30^\circ$ N/S  such as subtropical south Africa, eastern Brazil, north Australia, western India, eastern Siberia, eastern Mediterranean sea,  the western mountainous  regions of America (south-western North America, western Mexico, Andes) and the region between  Middle-east and the Hindu Kush-Karakoram mountain ranges.  Sub-Saharian Africa, West India  and the area in the Pacific Ocean  west of Ecuador have the highest global values ($\ge 1.4$).    The equatorial regions roughly located between $5^{\circ}$ S and $5^{\circ}$ N, where  convective rainfall is almost  permanent (Amazon and Congo basins, Indonesian Isles), have values of  $D$ less than 0.2.    Midlatitude regions as eastern U.S. and northwestern Europe also have low values of relative entropy ($\le 0.2$) because  baroclinic eddies deliver rain fairly constantly throughout the year.  Exceptions are the  eastern and southern Mediterranean coasts  and eastern Pacific, which instead are relatively dry during the boreal summer.  %Since $D$ tends to have large values where the annual precipitation distribution is very peaked  with a short wet season, maps of RE as those in Fig.~\ref{seconda}(c, d) are effectively maps of the inverse  length of the rainy season (Appendix A). 

\begin{figure*}
 \centering
\noindent\includegraphics[trim = 5mm  0mm 10mm  120mm,clip,angle=0, width=1.3\textwidth]{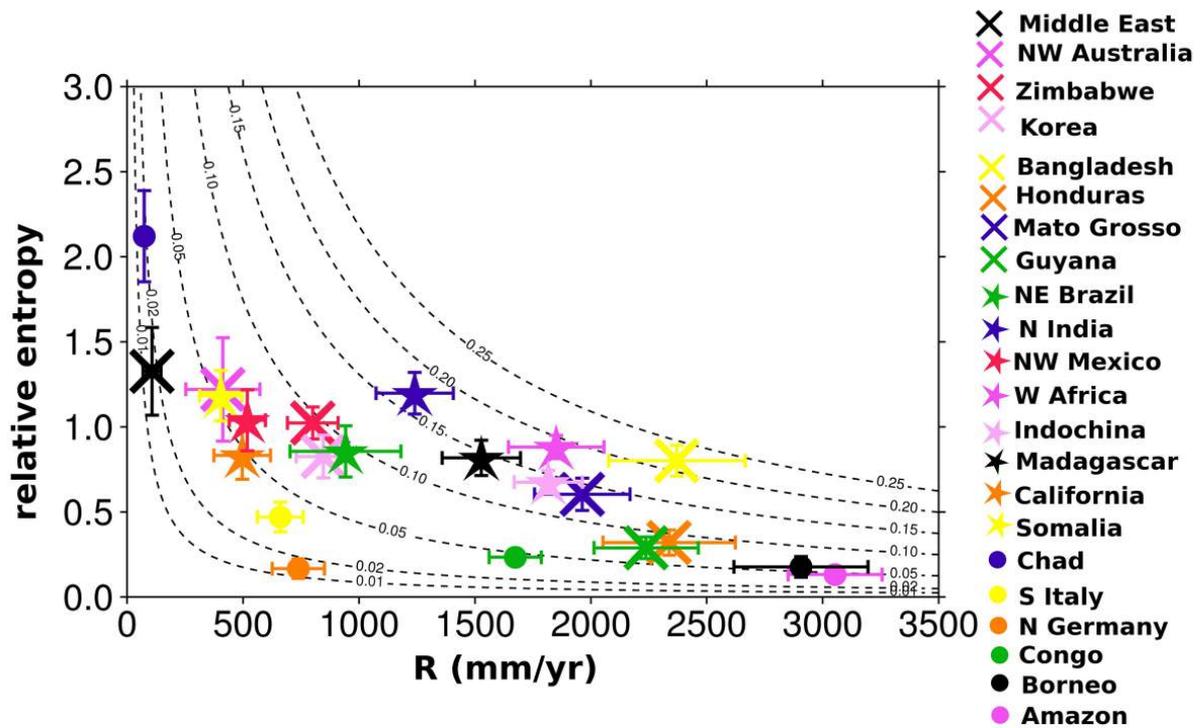} 
\caption{Precipitation-relative entropy diagram for different climatological areas for both GPCC dataset. Overplotted are isolines of dimensionless seasonality index.   Error bars denotes the range of inter-annual variability. Monsoonal precipitation regimes typically have  $S\ge 0.05$.  \label{sesta} }
\end{figure*}

In Fig. ~\ref{seconda}(e)  the DSI is shown over land    for the GPCC dataset while its patterns over oceans can be seen   in Fig.~\ref{seconda}(f) from  CMAP dataset.  Since $R$ and $D$ are generally observed to be  negatively correlated, the largest values of $S$ are generally  found in regions  with intermediate levels of  annual rainfall such as northeast region of Brazil, western  Africa, northern Australia and western Central America. The DSI  is   high also in parts of South and Southeast Asia where  both the total annual rainfall and RE  can be also extremely high (e.g. the Bengal region). % These regions of high seasonality are referred to as the global monsoon region \citep{Trenberth00, Ding}.  The DSI  is  a suitable tool for studying  precipitation regimes of monsoonal climates.
Equatorial regions (Indonesia, Congo basin, Amazon) and midlatitudes tend to have low values of $S$ because  their relative entropy is very small. Furthermore, although one order of magnitude smaller than typical values of  tropical regions, there is appreciable seasonality along the west coast of North America and Southwest U.S., midwest plains, Mediterranean regions, Middle-east, east Siberia and  Andes.   The  precipitation-relative entropy diagram of Fig.~\ref{sesta} provides an example of typical $R$, $D$ and $S$ values  averaged over  specific high-latitude, subtropical and tropical  areas.

 %coarse graining properties
 It is worth noting, at this point, an interesting  coarse graining property of $D$.  If we partition the year $Y$ in $N$  parts, we can define $D_N=\sum_1^N p_j \log_2(N p_j)$ with $p_j=r_j/R$ and $r_j$ the accumulated  precipitation over the $j$th time period  $Y/N$  (e.g. $N=12$ for  monthly precipitations).  In general $D_N\ne D_M$ for $N\ne M$ so the use of  different time bins (e.g. months and pentads) will give different values of relative entropy.  However, general properties of the RE allow us to say that if $N\ge M$, then $D_N \ge D_M$    and $S_N \ge S_M$  (see Appendix \ref{A3} for a formal proof). This  gives us confidence  on high-$D$ values,  which therefore   must be lower than   the ``true''   $D$. This general coarse-graining property is very useful to set lower bounds to  values of $D$ and $S$ at any chosen time resolution. 
%This property allowing a quantitative control of the indexes as time resolution is made coarser   makes $D$ superior to other possible indexes of seasonality.   
  In Fig.~\ref{pent}(a) we show  $D_{73}-D_{12}$   (pentads minus months)  and   note that the values of relative entropy obtained from pentad means  are  slightly larger than those  derived from monthly means. The error is less than $0.1$  over most of the global surface,  except in the  subtropical high regions where it amounts to about $0.2$-$0.4$.  Since these areas are very dry, the dimensionless seasonality index $S$ will not be  significantly  affected (Fig.~\ref{pent}(b)). We note  that regions featuring large values of $S$ (e.g. $S\ge 0.05$) have  errors $S_{73}-S_{12}$ generally smaller than $5\cdot 10^{-3}$. Since  $S$ is almost unaffected by the choice of the time resolution in regions where $S$ is large, as  in the global monsoon region \citep{Trenberth00, Ding}, it   is  a particularly robust index for studying  precipitation regimes of monsoonal climates.

 \subsection{Comparison between land datasets}

\begin{figure*}
 \centering
\subfigure{\includegraphics[angle=0, width=1.2\textwidth]{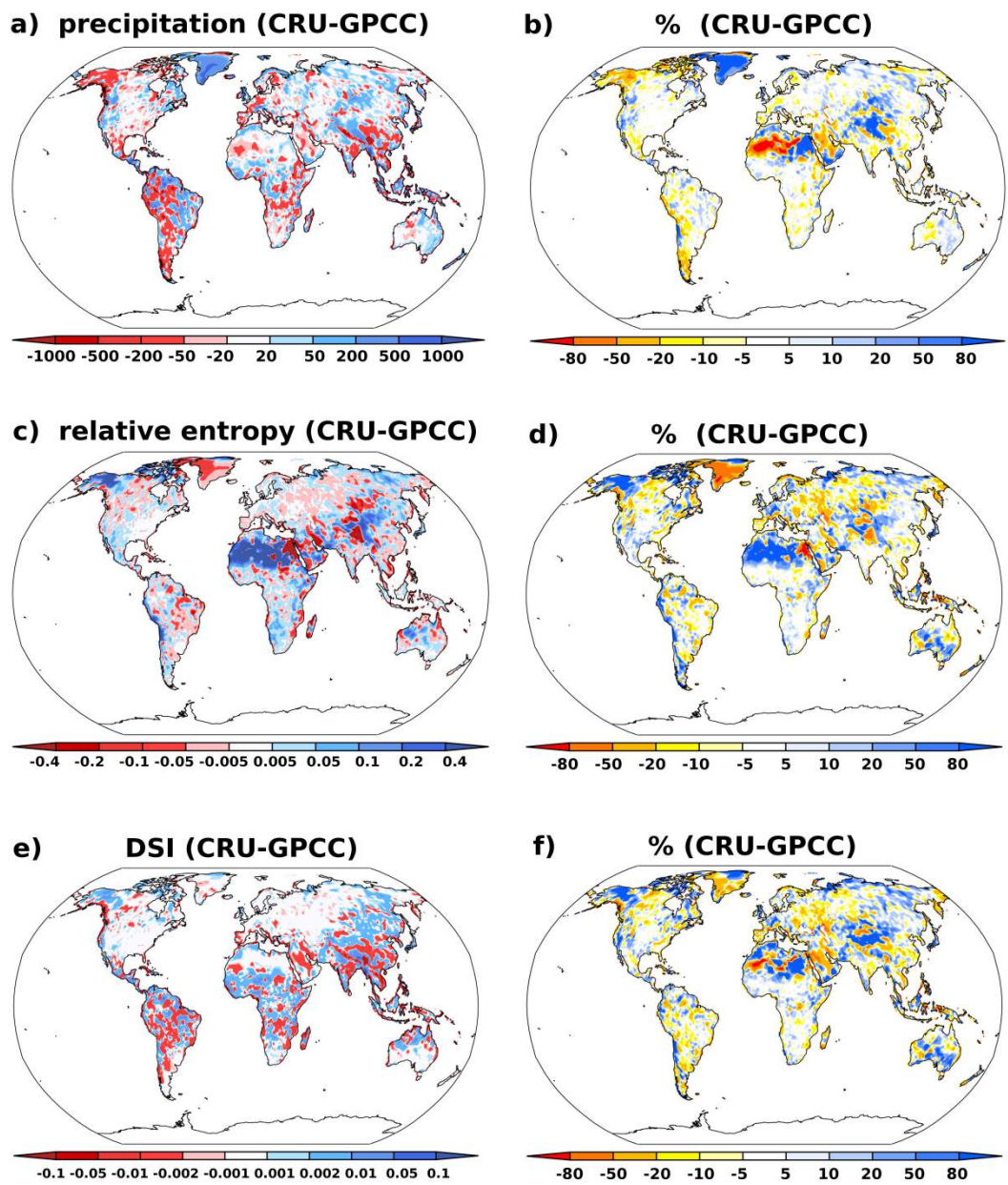} }
\caption{  Differences (a, c, e) and relative differences (b, d, f)  between the CRU and the GPCC datasets for the same quantities  in Fig.~\ref{seconda} over the period 1950-2010.   \label{terza} }
\end{figure*}

 Differences between the land-based CRU and GPCC datasets are shown in Fig.~\ref{terza}. Inconsistencies of the mean   annual  precipitation (Fig.~\ref{terza}(a, b))  in the two gridded datasets have already been documented in detail   by \cite{GPCC}; instead here  we  mostly focus on  the differences in the rainfall  seasonality   in terms of    $D$   (Fig.~\ref{terza}(c, d)). We note  relative differences of $D$  up to  $20\%$  in the semiarid or arid  regions  of Sahara, Middle-East and central Asia and also in areas where  the two datasets agree reasonably well in terms of annual total precipitation. 

As discussed in Appendix, $D$ is related to the number of wet months.    Therefore a difference $D_{\mathrm{cru}}-D_{\mathrm{gpcc}}$ implies a relative difference of the number of wet months  $n^\prime_{\mathrm{gpcc}}/n^\prime_{\mathrm{cru}}\approx 2^{(D_{\mathrm{cru}}-D_{\mathrm{gpcc}})}\approx 0.3$ for $D_{\mathrm{cru}}-D_{\mathrm{gpcc}}=0.4$ and $n^\prime_{\mathrm{gpcc}}/n^\prime_{\mathrm{cru}}\approx 0.15$ for  $D_{\mathrm{cru}}-D_{\mathrm{gpcc}}=0.2$. Considering that in such semiarid regions (e.g. sub-Saharan Sahel) precipitations are concentrated within one-two months, such  uncertainties in $D$ reveal fairly large inconsistencies between the two observational datasets in reproducing the time distribution of the precipitation events. On the contrary, in regions such as the slopes of the Himalaya, where there are large differences in the annual total rainfall (up to $1000$ mm/year), differences in relative entropy are relatively small ($\approx 0.05$) and hence  the two datasets agree reasonably well in reproducing the monthly precipitation signal.
Differences in the DSI  between CRU and GPCC  are shown  in Fig.~\ref{terza}(e, f). The two land datasets show  remarkable differences in the seasonality index  over southern and central  America, most of Africa and south-southeastern Asia.  Such inconsistencies in the observational datasets are due  to differences in the annual total precipitation in the case of  South America,  in the relative entropy   for  sub-Saharan Africa and central Asia,  and to both terms in the case of east Indochina and Madagascar.

\subsection{The dimensionless seasonality index and the global monsoon regions}
\label{dom}

\begin{figure*}
 \centering
\subfigure{\includegraphics[trim = 4mm  70mm 0mm  70mm,clip,angle=-0, width=0.9\textwidth]{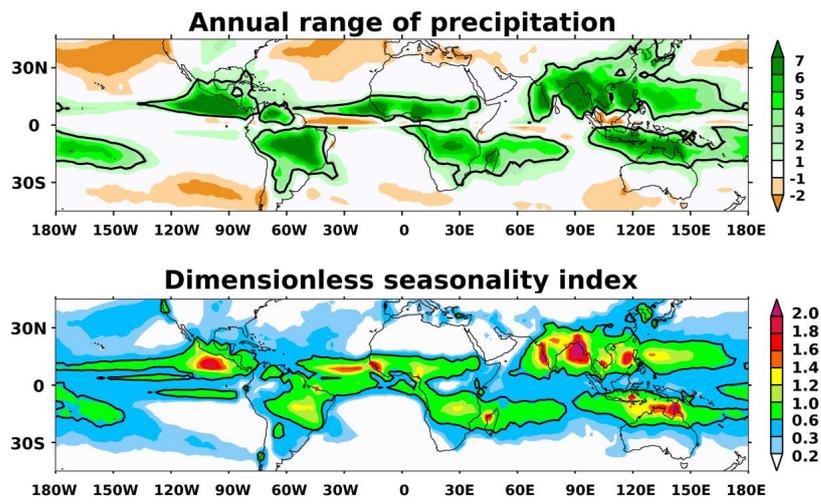} }
\caption{ Annual range of precipitation  (mm/day) and DSI ($\times 10$) for the CMAP climatology. The global monsoon domain (\emph{black thick line}) is defined by the annual range equal to   $2.5$ mm/day and DSI equal to  $0.05$.   \label{domain}}
\end{figure*}

The DSI combines information about seasonality  and intensity of rainfall and therefore it is  a useful indicator of the extent of monsoonal precipitation regions \citep{Ding}. Rainfall is indeed  the most important monsoon variable given the high socioeconomic and ecological impact, and indexes based on rainfall are widely used to study the global monsoon \citep{ Wang, Kitoh, LeeWang}.
In Fig.~\ref{domain} the DSI is compared to the Annual Range of Precipitation (ARP)   \citep{Ding, Wang}. The  ARP  is defined as the   local summer  minus winter precipitation rate, i.e.     the   MJJAS minus NDJFM precipitation rate in the northern hemisphere and  NDJFM minus  MJJAS in the southern hemisphere. %  \cite{Wang} define  the domain of the global monsoon as the region where  the ARP exceeds  $2.5$ mm\,day$^{-1}$.  

Direct comparison of the the two indicators in Fig.~\ref{domain} shows that regions featuring  high DSI    capture  fairly well the global monsoon region.  The global   monsoon domain,     defined   by  \cite{Wang} as the area in which   the ARP is greater than $2.5$ mm\,day$^{-1}$, is described also    by the isoline   $S \approx 0.05$.  The two domains match pretty well over land in  Africa,   Central-South  America and Australia although differences are found over the Atlantic  and Pacific ocean.   Let us note that  the DSI is positive definite     whereas the ARP has negative values outside the Tropics,  where midlatitude precipitations occur during the local winter. While differences between the two criteria appear to be minor, they may still be relevant for assessing the robustness of future changes  of the global monsoon domain \citep{Kitoh, LeeWang}.   Shifts of the borders of the monsoons domain may be especially critical  for  areas located at the   border of monsoonal circulation, as  for example the Indus basin \citep[e.g.][]{Shabeh2} or the arid North America Southwest \citep{Cook}. These areas   might go through critical  changes in their precipitation regimes if the extent  of the  rainfall associated with the monsoonal circulation shifts aways or it is delayed \citep{Seth}.        \cite{Kitoh} show that under RCP8.5 scenarios the global monsoon areas as defined by the ARP  is mostly unaffected,  with  little changes over  central and  eastern tropical Pacific, eastern Asia and southern Indian ocean. Given the small entity of such changes, we stress here the importance of different and alternative monsoon indexes  to assess the robustness of  future changes in monsoonal precipitation  and give more confidence to results on changes of the global monsoon.

\section{Comparison with CMIP5 coupled climate models}
\label{intercomp}

In this section we evaluate the mean total annual precipitation and the relative entropy  for the GCMs listed in Table~\ref{tab1} and compare it with the same indicators estimated for observational datasets. The aim is to assess models' skill in reproducing the rainfall seasonality as defined by $D$ and show the use of RE maps  to determine areas of interest that  need  more detailed analyses of seasonality  using more traditional methods.

\subsection{Annual precipitation and relative entropy}

\begin{figure*}
 \centering
\subfigure{\includegraphics[trim = 5mm  0mm 0mm  70mm,clip,angle=-0, width=1.5\textwidth]{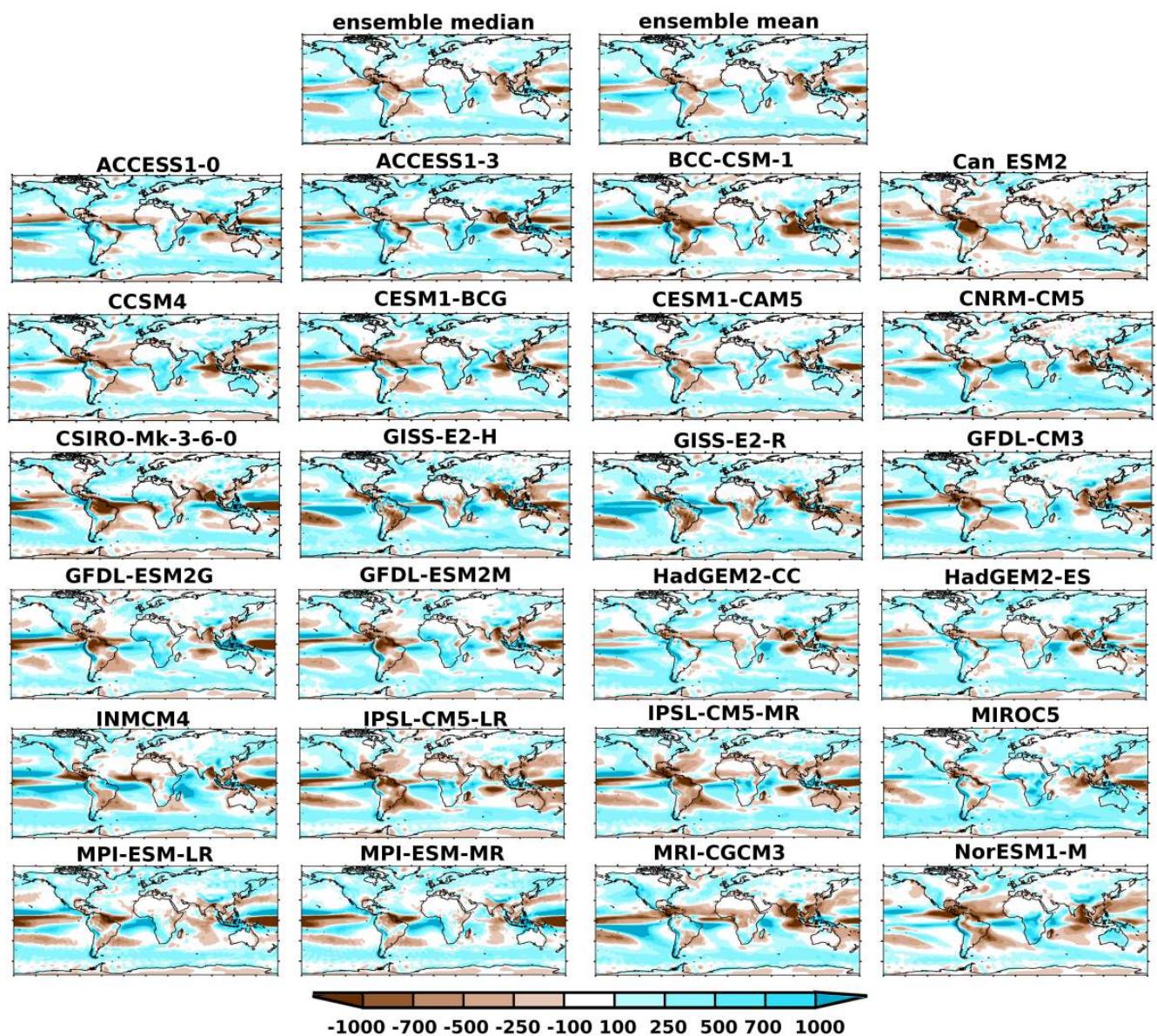} }
\caption{ Biases of the CMIP5 models  mean precipitation (mm)  over the period   1979-2009  relative to the CMAP climatology. The multimodel ensemble median and mean are also shown. \label{prec_all}}
\end{figure*}

\begin{figure*}
 \centering
\subfigure{\includegraphics[angle=0, width=0.9\textwidth]{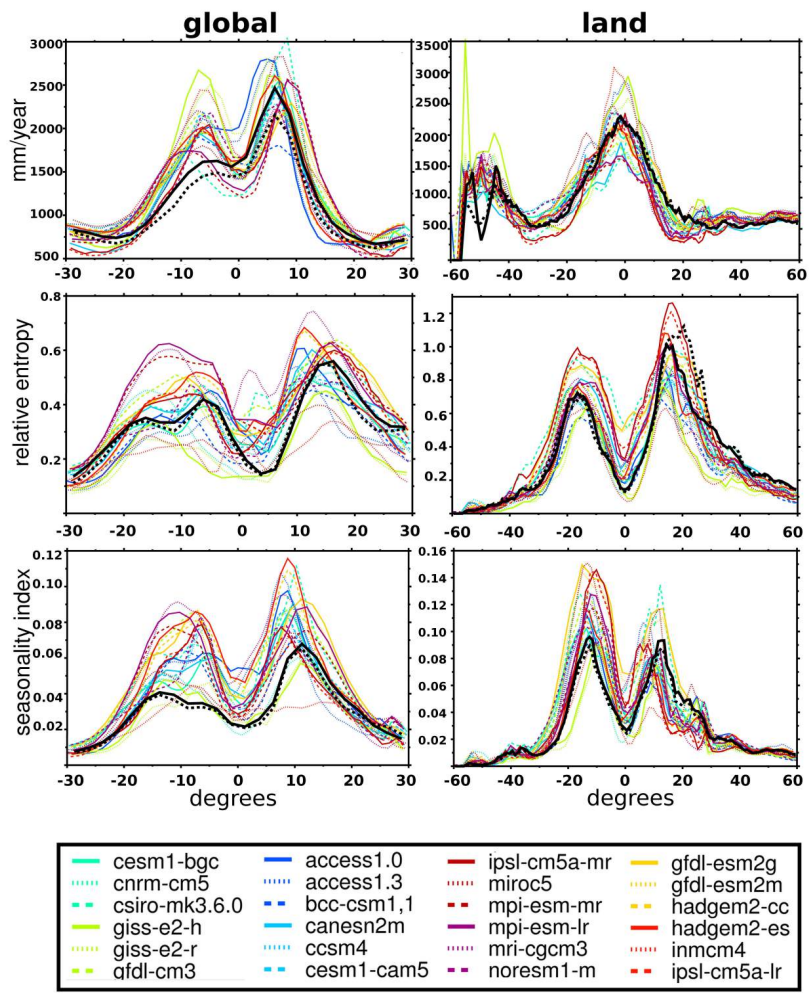} }
\caption{Left column: zonal means of precipitation, relative entropy and dimensionless seasonality index over the whole globe (GPCP and CMAP observation  dataset are denoted with black continuous and dashed line respectively) for the period 1979-2008. Right column: as before but over land only (GPCC and CRU are denoted with black continuous and dashed line respectively). \label{intercomparison}}
\end{figure*}

Mean annual rainfall differences between the CMIP5 models listed in Table~\ref{tab1}   and CMAP observations are     shown in Fig.~\ref{prec_all}.  It is noted that  the  double ITCZ problem \citep{Lin} -- the additional band of precipitation south of the equator in the Pacific ocean --   affects most of the CMIP5   models  and it is  particularly  strong   in those  with low resolution such  as   the  GISS models or the INMCM4.  The double ITCZ bias  is one of the most persisting GCMs bias and there has been little improvement from CMIP3  to CMIP5 models \citep{Hwang}. In CMIP5 models it improves   as the model resolution increases (MIROC5), although it persists also in models with high horizontal resolution (MRI-CGCM3). The zonal mean of the global precipitation field (Fig.~\ref{intercomparison}) shows  the excess of rainfall due to the double ITCZ. In Fig.~\ref{intercomparison} it can also be seen that  models generally show a large spread in the latitudinal position of the maxima of  zonal mean of precipitation.  These problems may be   directly related to how models simulate the  ocean meridional heat transport \citep{Frierson}. The ITCZ problems seems therefore to be  constrained by the surface heat fluxes and  thus related to the capability of models to correctly  simulate clouds and other controls of  the surface solar energy flux \citep{Hwang}.  

 Overall,  CMIP5 models tend to have a too large RE over  tropical Latin America and a too small RE over Western Africa, Western Mexico and East Asia. This is evident from the multimodel ensemble mean MME and median MMM (Fig.~\ref{entr_all}).  Most of models reproduce fairly well the RE pattern over Southern Africa, with negative biases exhibited only by few models (BCC-CSM-1, GISS-E2-R, GISS-E2-H). The MME and MMM feature  small biases also over Australia, due to the cancellation of large positive (e.g. IPSL-CM5-LR) or negative (e.g. GISS-E2-R) biases  shown by single models.    A   large  negative RE bias  over East Asia, extending  from north-eastern China up to the Tibetan region,    is present in all models and it is particularly severe in some GCMs (up to $-0.5$, e.g. in GISS-E2-R).  The inmcm4 and the GISS models have a general tendency to underestimate  RE over both land and oceans, whereas the MRI-CGCM3 and MPI-ESM models tend to overestimate  RE.

 %a  wide subtropical   region ranging from Western Sahara  through the Arabian Peninsula up to the Indus basin and over central and southern Australia.  Both these  regions  are characterized by arid and semiarid climates (precipitation typically $\le 100$ mm/year) placed at the borders of the Western African Monsoon regions (Sahara) and the Australian monsoon region (central-southern Australia). Furthermore,  most of  models also   show a negative bias in the RE over the Western African  and the East Asian monsoon regions. }  % According to the interpretation of $D$ given in   Appendix \ref{appendix},    a positive bias  suggests  that  a too peaked rainfall distribution  (shorter duration of the wet season)  and a negative bias suggest that  a too flat rainfall distribution (longer duration of the wet season)  throughout the year  is simulated by CMIP5 models.  
 
Zonal means (Fig.~\ref{intercomparison}) reveal that  models generally perform better over land.  However they show a large  dry bias in the precipitation at around $20^\circ$ N associated with the  South Asian Monsoon \citep{Turner, Sperber, Shabeh1, Shabeh2,Boos} over the Indian region (e.g.  HadGEM2, Fig.~\ref{prec_all}).  Central America and northern South America also feature strong negative rainfall biases  \citep{HwangSupp}. 
 MPI-ESM-LR  largely  overestimates the RE   in the Southern Hemisphere, although it behaves fairly realistically in the Northern one (Fig.~\ref{intercomparison}). In the equatorial zone ($10^\circ$ S-$10^\circ$ N) we  observe  that most of the CMIP5 model feature  very large values of RE (for example over the Amazon and eastern Africa  region, with GFDL-ESM2G, GFDL-ESM2M or CSIRO-Mk3.6.0 having $D\ge 0.4$, error of order $400 \%$) in areas characterized by   values of RE typically smaller than 0.2.
 %This implies that   models   tend to simulate an unrealistic seasonal cycle for rainfall in the deep  tropics.   
  Exceptions are the  GISS-E2-H, GISS-E2-R, MIROC5 and INMCM4 which instead underestimate  the RE over the tropics.   Overall CESM1-CAM5 model  is the best at simulating the observed RE.

Let us note, to conclude this biases analysis, that models errors in RE are not removed by a simple mean bias adjustment of $r_m$.   Bias adjustment algorithms have been developed to bring GCMs simulations closer to observations before applying statistical and dynamical downscaling \citep[e.g.][]{Christensen, Li}.
 Given the  monthly precipitations $r_{m,j}$ (month $m$, year $j$) and the observations $\rho_{m,j}$,   a mean bias adjustment would lead to new monthly precipitations $r_{m,j}^\prime = \alpha_j r_{m,j}$, with $\alpha_j = \sum_m \rho_{m,j}/ \sum_m r_{m,j}$ and   bias adjusted rainfall fractions $p^\prime_{m,j}=$$ r^\prime_{m,j}/\sum_m r^\prime_{m,j}=$ $ \alpha_j r_{m,j}/\sum_m  \alpha_j r_{m,j}=$$p_{m,j}$. Thus,  a  mean bias adjustment does not affect the monthly precipitation fractions, thus leaving  the  relative entropy unaltered.

\begin{table}
\caption{  List of the CMIP5 models used for this study.  Numbers are used to identify them in Fig.~\ref{taylor} \label{tab1}}
\begin{center}
\footnotesize
\resizebox{\columnwidth}{!}{% 
\begin{tabular}{ l l l l c}
\hline
\toprule
Number & Model  name      & Modelling Centre  & Country & AGCM resolution (lon$\times$lat) \\ %& OGCM resolution \\
\hline
1& ACCESS1.0       & CAWCR $^\textrm{\textbf{a}}$&Australia                   & 192$\times$145/L38 \\% & 360$\times$300/L50\\
2&  ACCESS1.3                   & CAWCR &Australia      &   192$\times$145/L38  \\%& 360$\times$300/L50 \\
3 & BCC-CSM1.1   & BCC $^\textrm{\textbf{b}}$&China                                &128$\times$64/L26  \\%&  1$^\circ\times$1.33$^\circ$/L40 \\
4&  CanESM2         & CCCMA $^\textrm{\textbf{c}}$ & Canada                           & 128$\times$64/L35                        \\%& 256$\times$192/L40 \\
5& CCSM4                        &  NCAR $^\textrm{\textbf{d}}$ &USA               &  288$\times$192/L26  \\%&1.11$^\circ\times$0.27$^\circ$--0.54$^\circ$/L60\\
 6& CESM1-BGC               &NCAR&USA                  & 288$\times$192/L26  \\% & 1$^\circ\times$1$^\circ$/L60 \\ 
7 & CESM1-CAM5                   & NCAR&USA          & 288$\times$192/L30  \\%& 1.11$^\circ\times$0.27$^\circ$--0.54$^\circ$/L60 \\
 8& CNRM-CM5      &  CNRM/CERFACS $^\textrm{\textbf{e}}$&France            &     256$\times$128/L31  \\% & 1$^\circ\times$1$^\circ$/L42\\
 9& CSIRO-Mk3.6.0    &   CSIRO/QCCCE $^\textrm{\textbf{f}}$& Australia       & 192$\times$96/L18    \\%&1.875$^\circ\times \sim 0.9375 ^\circ$/L31\\
 10& GISS-E2-H        &  GISS $^\textrm{\textbf{g}}$& USA                                  &  144$\times$90/L40   \\%&  1.25$^\circ\times$1$^\circ$/L32\\
  11& GISS-E2-R        &  GISS & USA                                  &  144$\times$90/L40   \\%&  1$^\circ\times\sim$1$^\circ$/L32\\
 12& GFDL-CM3                           &GFDL $^\textrm{\textbf{h}}$&USA             &144$\times$90/L48    \\% &360$\times$200/L50\\
 13& GFDL-ESM2G                  &GFDL&USA                &144$\times$90/L24     \\%&360$\times$210/L63\\
 14& GFDL-ESM2M                   & GFDL&USA             &144$\times$90/L24     \\%&360$\times$200/L50\\
 15& HadGEM2-CC    &MOHC $^\textrm{\textbf{i}}$&UK                                   &  192$\times$145/L60          \\% &   1.875$^\circ\times$1.25  $^\circ$/L40   \\
 16& HadGEM2-ES    &MOHC &UK                                   &  192$\times$145/L38     \\%  &     1.875$^\circ\times$1.25  $^\circ$/L40   \\
 17& INMCM4             & INM $^\textrm{\textbf{j}}$&Russia                                & 180$\times$120/L21     \\%&  1$^\circ\times$0.5  $^\circ$/L40 \\
 18& IPSL-CM5A-LR               & IPSL $^\textrm{\textbf{k}}$&France            &96$\times$95/L39    \\%&  2$^\circ\times$2$^\circ$/L31\\
 19& IPSL-CM5A-MR              &IPSL &France             &96$\times$95/L19     \\%& 2$^\circ\times$2$^\circ$/L31\\
20&  MIROC5            & MIROC $^\textrm{\textbf{l}}$ &Japan                            &256$\times$128/L40     \\%&    256$\times$224/L50 \\
21& MPI-ESM-MR    &MPI-M $^\textrm{\textbf{m}}$ &Germany                        &192$\times$96/L95    \\%&  1.5$^\circ\times$1.5/L40  \\    
22& MPI-ESM-LR     & MPI-M &Germany                      &192$\times$96/L47      \\%& 1.5$^\circ\times$1.5/L40  \\
23& MRI-CGCM3     & MRI $^\textrm{\textbf{n}}$ & Japan                             &  $320\times 160$/L48   \\%& 1$^\circ\times$0.5$^\circ$/L51 \\
24& NorESM1-M      & NCC $^\textrm{\textbf{o}}$ &Norway                           & 144$\times$96/L26   \\%& 384$\times$320/L53 \\ 
% & FGOALS-g2               &LASG/IAP&China           &128$\times$60/L26 & 360$\times$196/L30\\
% & FGOALS-s2                 &LASG/IAP&China         &128$\times$108/L26 & 0.5 $^\circ$--1$^\circ\times$0.5 $^\circ$-- 1$^\circ$  \\
\bottomrule
 \hline
 \multicolumn{5}{l}{%
  \begin{minipage}{15cm}%
    \tiny \quad \\ \\\textbf{a} Centre for Australian Weather and Climate Research; \textbf{b}  Beijing Climate Centre, China Meteorological Administration; \textbf{c} Canadian Centre for Climate Modelling and Analysis; \textbf{d} National Center for Atmospheric Research; \textbf{e} Centre National de Recherchers Meteorologiques/Centre Europeen de Recherche et Formation Avancees en Calcul Scientifique; \textbf{f}  Commonwealth Scientific and Industrial Research Organization/Queensland Climate Change Centre of Excellence; \textbf{g} NASA Goddard Institute for Space Studies; \textbf{h} NOAA Geophysical Fluid Dynamics Laboratory; \textbf{i} Met Office Hadley Centre; \textbf{j} Institute for Numerical Mathematics; \textbf{k} Institute Pierre-Simon Laplace; \textbf{l}  Atmosphere and Ocean Research Institute (The University of Tokyo), National Institute for Environmental Studies, and Japan Agency for Marine-Earth Science and Technology; \textbf{m} Max Planck Institute f\"ur Meteorologie; \textbf{n}  Meteorological Research Institute  ; \textbf{o} Norwegian Climate Centre%
  \end{minipage}%
}\\
\end{tabular}
}
\end{center}
\end{table}

\begin{table}[htdp]
\caption{  Coordinates boundaries (east, south, west, north) of the rectangular areas shown in Fig.~\ref{sesta} (upper part of the table) and Fig.~\ref{entr_bias}-Fig.~\ref{SAM} (lower part of the table). \label{tab2} }
\begin{center}
\begin{tabular}{ l  r  l   r   r   r}
\hline
\toprule
Area & East       & South & West & North \\
\hline
Mato Grosso     &      $60^\circ$ E   &  $14^\circ$ S &   $55^\circ$ W       &  $10^\circ$ S   \\
NW Australia        &        $125^\circ$ E     & $23^\circ$ S     &  $135^\circ$ E  & $18^\circ$ S  \\
N India&   $75^\circ$ E  &   $20^\circ$ N      &   $88^\circ$  E       & $25^\circ$ N   \\
NE Brazil  &  $45^\circ$ E    &      $10^\circ$  S    &    $35^\circ$  E         &  $0^\circ$  N\\
Somalia     & $40$ E& $0$ N & $50$ E &  $10$ N\\ 
 NW Mexico    &$112$ W & $20$ N & $104$ W & $32$ N \\
 S Italy    & $12$ E & $36$ N&$19$ E & $42$ N\\ 
 Congo     & $15$ E & $5$ S& $25$ E& $5$ N\\
 N Germany    & $7$ W &$51$ N &$12$ E & $55 $ N\\  
 Chad    & $15$ W & $15$ N  & $23$ W  & $20$ N\\ 
 Borneo    & $108$ E & $4$ S & $119$ E  & $8$ N\\ 
 Amazon    & $70$ W&$5$ S & $65$ W &  $5$ N\\
 Indochina    & $100$ E & $10$ N & $110$ E & $20$ N \\ 
 Madagascar    & $42$ E & $25$ S & $50$ E & $12$ S\\ 
 Brazil/Guyana    & $65$ W & $3$ S & $55$  W& $3$ N \\
 California    & $125$ W &  $33$ N  & $115$ W  &  $42$ N\\ 
 Guinea    & $20$ E  & $0$ N  & $8$ E  & $14$ N \\ 
 Korea    &$115$ E &  $35$ S & $130$ E &  $41$ N\\  
 Bangladesh    &$88$ E   & $22$ S &$92$ E & $26$ N \\
  Zimbabwe    & $20$ E  & $25$ S  & $35$ E &  $10$ S\\ 
 Honduras    &  $86$ W & $13.5$ N & $84$ W & $15.5$ N \\
 Middle East    & $36$ E& $30$ N & $40$ E & $34$ N \\ 
 \bottomrule
 Australia & 120 E & 30 S& 150 E & 20 S \\
 Sub-Sahara   & 10 W & 15 N & 30 E & 23 N\\
 West Africa & 0 E&5 N &30 E & 15 N \\
 Tropical South America & 60 W & 20 S & 40 W & 5 S \\
 East Asia & 90 E & 30 N & 120 E & 50 S \\
  \bottomrule
\hline
\end{tabular}
\end{center}
\end{table}

\begin{table}
\caption{  Correlation pattern (PC), root mean square error (RMSE) and standard deviation ($\sigma$)  of      the observational datasets (GPCC, CRU, GPCP, CMAP) and of        the CMIP5 models (Fig.~\ref{taylor}).  Comparison is made with the GPCC dataset and restricted over land  for the 1950-2010 climatology.  Values of RMSE and $\sigma$ are normalized with respect to the standard deviation of  GPCC $\sigma_{\textrm{gpcc,R}}=807$\,mm, $\sigma_{\textrm{gpcc,D}}=0.47$ and $\sigma_{\textrm{gpcc,S}}= 0.034$. For observations, bold character    highlights values of PC and $\sigma$ defining   the observation  uncertainty range.    Models which are  placed within the  observation uncertainty range   are  also highlighted  in bold  whereas those farthest  from the reference  (the largest RMSE) are highlighted in bold italics.  \label{tab3}}
\begin{center}
\footnotesize
\resizebox{\columnwidth}{!}{% 
\begin{tabular}{ l l l c c c l c c c l c  c c}
\hline
\toprule
number/ & model & &\multicolumn{3}{c}{annual precipitation}& &\multicolumn{3}{c}{relative entropy}& &\multicolumn{3}{c}{DSI }\\
\cline{4-6}
\cline{8-10}
\cline{12-14}
color &    & & PC  & RMSE & $\sigma$& & PC  & RMSE & $\sigma$&  &PC  & RMSE& $\sigma$ \\
\hline
\toprule
reference    & GPCC              & &\bf 1    &  0   & $807$\,mm &                             &\bf 1&0 &0.47  &                & \bf 1&0& 0.034\\
 black   & CRU                 &&\bf 0.83& 0.57 &  0.97 &                                & 0.79&0.60  & 0.74&            & \bf 0.92& 0.39 &\bf 1.07 \\
 green   & GPCP              & &0.94& 0.32 & 0.89 &                          & \bf 0.72& 0.69&\bf 0.63 &            & 0.95  & 0.31  & 1.03  \\
  magenta  & CMAP             & &0.95&  0.32  & \bf 0.84&                     & 0.78&0.63 &0.67 &            & 0.93  & 0.35 & \bf 0.91  \\
    \hline
    
1  & ACCESS1.0   & &\bf 0.84&  0.54 &\bf 0.87 &                     & 0.62& 0.78& 0.57&                      &0.77 &0.66&0.97 \\
2  & ACCESS1.3   & &0.82 & 0.62  & 1.08  &                      &0.59 &0.80 &0.58&                 &0.74&0.84&1.24  \\
3  & BCC-CSM1.1   & &0.78&0.62 &  0.75 &                                &0.63&0.80&0.44 &                 & 0.71& 0.74&  0.92\\
4  & CanESM2       & &0.76&\bf  \textit{0.65} &0.68 &    &\bf0.73&  0.68&\bf 0.71 &               &0.76 &0.71&1.06\\
5  & CCSM4         & &\bf 0.83& 0.55 & \bf 0.84&                   &0.71&0.71 & 0.58 &            &  0.77&0.76 & 1.18\\
6  & CESM1-BGC &&\bf 0.83& 0.55& \bf 0.84&                     &  0.72&0.70 &0.58&              &0.77&0.76 &1.18 \\ 
7  & CESM1-CAM5 &&0.83& 0.55  & 0.79&                   &\bf 0.73& 0.69 & \bf 0.63 &              &0.77& 0.72 & 1.13\\
8  & CNRM-CM5  &&0.83&0.57 &0.73 &                           &0.76&  0.67&0.57 &               &0.81 & 0.59&   0.89\\
9  & CSIRO-Mk3.6.0 &&0.75&\bf \textit{0.66}&0.81&   &0.62 & 0.80& 0.75&               & 0.72 &\bf \textit{0.98}&1.40\\
10& GISS-E2-H  &&0.75& \bf \textit{0.70}&0.99&     &0.57&\bf \textit{0.82}&0.49&    &0.65& 0.87&1.10\\
11& GISS-E2-R  &&0.78&   0.63& 0.92&                            & 0.54&\bf \textit{0.84}& 0.46 &               &0.67& 0.79& 0.96\\
12& GFDL-CM3 &&0.81&0.62   & 0.61 &                              & 0.70 & 0.71&  0.63  &            &0.77 & 0.68& 1.03\\
13& GFDL-ESM2G &&0.75& \bf \textit{0.65}&0.75& & 0.64 &0.77 &0.74&                 &0.77& \bf \textit{0.95} & 1.48 \\
14& GFDL-ESM2M &&0.76&  \bf \textit{0.64}& 0.71 &          & 0.65 &0.76 & 0.73 &                  & 0.76 &0.90&  1.39\\
15& HadGEM2-CC && 0.84&  0.53 & 0.82 &                 &0.63& 0.77& 0.66&                    & 0.76& 0.67 & 0.93 \\
16& HadGEM2-ES && 0.85 &  0.52   &0.83  &              &0.62 & 0.78&0.67 &               & 0.76& 0.67 &  0.97 \\
17& INMCM4         &&0.80&     0.59  & 0.91 &                     &0.57&\bf \textit{0.82} &0.83 &                  & 0.71& 0.70& 0.83\\
18& IPSL-CM5A-LR &&0.75&  \bf \textit{0.65}&0.71&  &0.64 &0.79 & 0.91 &                 & 0.64& \bf \textit{0.95}& 1.22\\
19& IPSL-CM5A-MR && 0.75& \bf \textit{0.65}&0.79&   & 0.68 &0.76 & 0.54&                  &0.66& \bf \textit{0.97}&1.29\\
20& MIROC5           &&0.80&  0.61   & 0.94 &                      &0.61 & 0.79&  0.70 &                 &  0.80& 0.87& 1.43\\
21& MPI-ESM-MR   &&0.82&  0.58  &0.70  &                         &   0.70& 0.70&0.68 &                   &  0.81&0.65 &1.10\\   
22& MPI-ESM-LR    &&0.83&   0.57  & 0.72  &                       & 0.67& 0.73&  0.64 &                  &  0.82&0.65&1.10 \\
23& MRI-CGCM3    &&0.81&   0.60   &  0.92  &                      &0.58 & \bf \textit{0.81}&0.60&                       & 0.70& 0.82& 1.10\\
24& NorESM1-M     &&0.77 & 0.64    & 0.89 &                      &  0.68 & 0.73& 0.48 &                    & 0.71&0.82&0.92\\
  blue  & MME              & &\bf 0.88&  0.47 & \bf 0.89  &                   & 0.73& 0.72 &0.48 &               & 0.85  & 0.52 & 0.92  \\
 \bottomrule
\end{tabular}
}
\end{center}
\end{table}

\subsection{Interpretation of RE  biases}

In order to clarify the reasons of the RE biases documented in the previous section, we focus on five areas --   tropical Latin America, central Australia, Sub-Saharan Africa,  Western Africa and East Asia  -- where either  most of the  models show consistent  biases  or some of them feature very large RE biases -- and analyze the origin of such biases in terms of precipitation fractions. Coordinates of their rectangular domains 
%over which the monthly rainfall fractions are area-averaged,  
are listed in Tab.~\ref{tab2}.

\begin{figure*}
 \centering
\subfigure{\includegraphics[trim = 8mm  0mm 0mm  75mm,clip,angle=-0, width=1.6\textwidth]{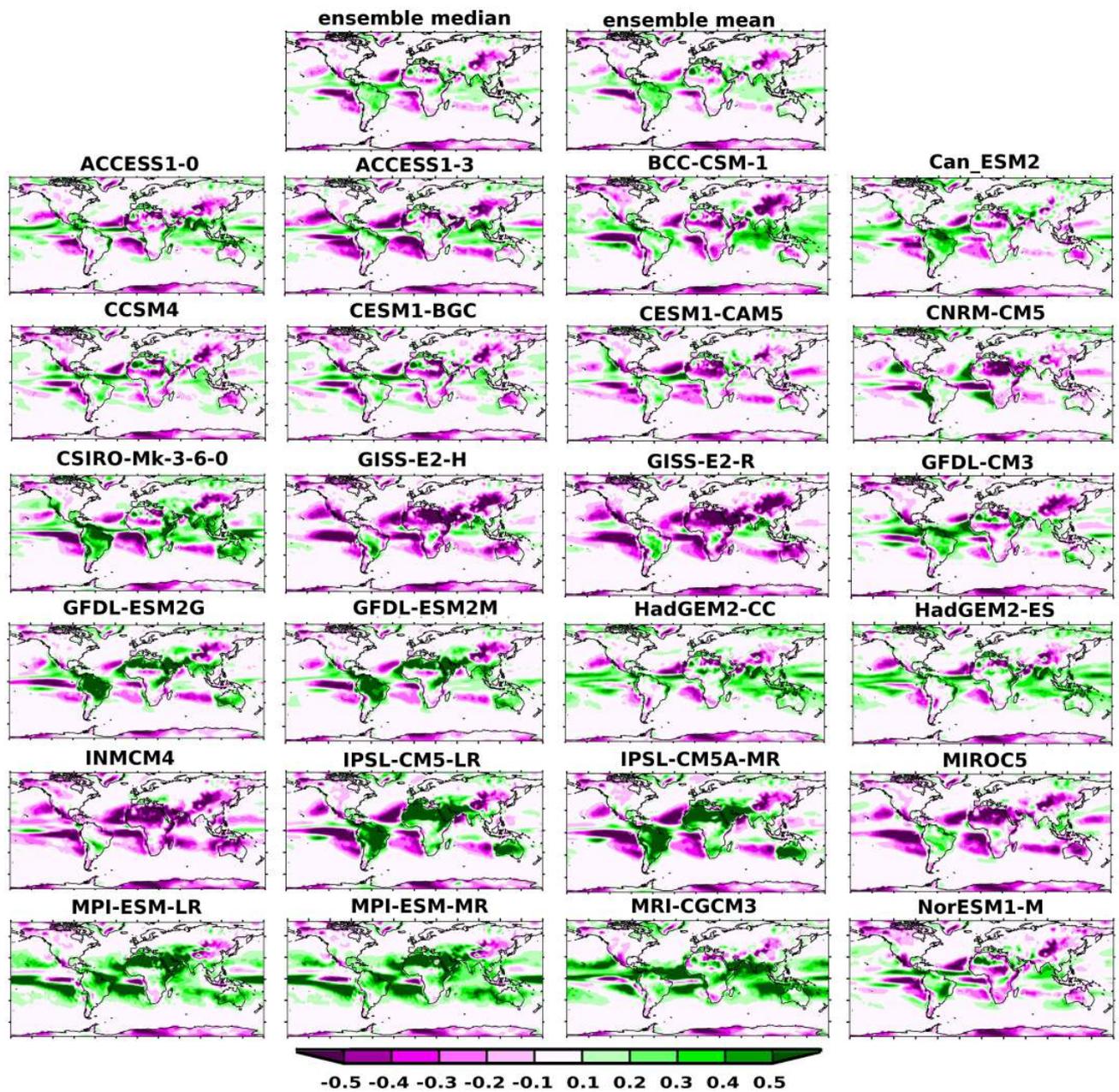} }
\caption{  As in Fig.~\ref{prec_all}, but for relative entropy.   \label{entr_all}}
\end{figure*}

\begin{figure*}[t]
\centering
  \noindent\includegraphics[trim = 0mm  90mm 0mm  30mm,clip,width=0.8\textwidth,angle=-0]{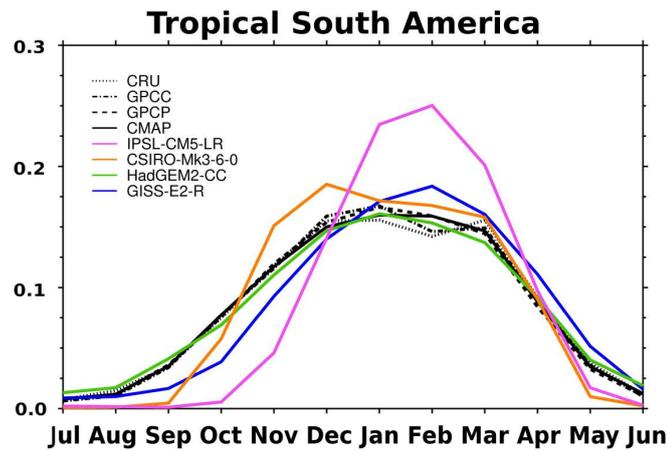}
  \caption{ Precipitation frequencies $p_m=r_m/R$ for observation datasets CRU, GPCC, GPCP, CMAP and four models (CSIRO-Mk3-6-0, IPSL-CM5-LR, HadGEM-CC and GISS-E2-R)  over tropical Latin  America. }\label{SAM}
\end{figure*}

Most of models show  a consistent positive RE biases  over tropical South America  (Fig.~\ref{entr_all}).  In  Fig.~\ref{SAM}  a comparison with observations  is shown for some of the most (IPSL-CM5-LR and CSIRO-Mk3-6-0)   and least (HadGEM2-CC and GISS-E2-R) biased CMIP5 models. The reason of the RE positive bias is particularly evident  from the IPSL simulation, which  exaggerates the December-March $p_m$ while underestimating them in the pre-monsoonal season. A similar behavior, though less accentuated, is observed also in the CSIRO model. A less severe  bias characterizes the  GISS model while the  HadGEM2-CC captures  extremely  well the monthly rainfall fractions and presents      almost no RE biases over the region.

\begin{figure*}[t]
\centering
  \noindent\includegraphics[trim = 0mm  110mm 0mm  0mm,clip,width=1.0\textwidth,angle=-0]{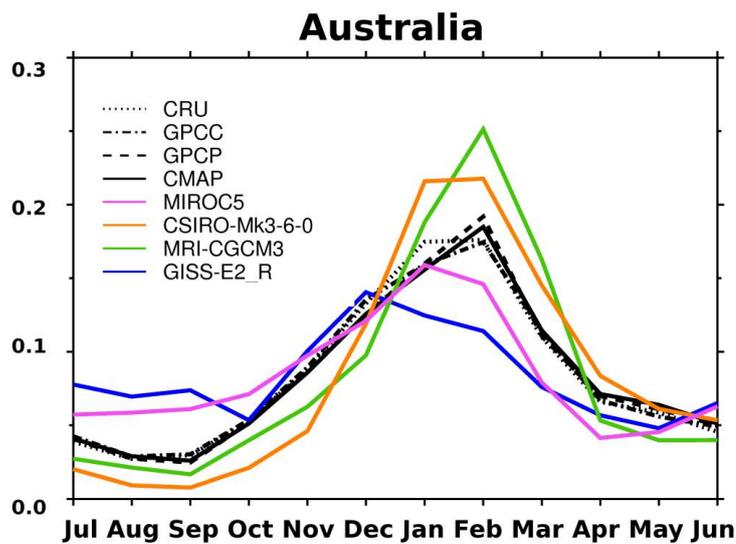}
  \caption{  Precipitation fractions $p_m=r_m/R$ for observation datasets CRU, GPCC, GPCP, CMAP and four models (CSIRO-Mk3-6-0, MRI-CGCM3, MIROC5 and GISS-E2-R) over Australia.}\label{entr_bias}
\end{figure*}

In Fig.~\ref{entr_bias}  monthly precipitation frequencies $p_m$  are shown for observations  and   four CMIP5 models (MRI-CGCM3,  CSIRO-Mk3-6-0, MIROC5 and GISS-E2-R) over  the Australian region. The first two models   show  positive biases in RE whereas the last two models  have   negative biases.    Note that the mean annual rainfall does not show large biases in these regions, so the anomalies in  RE must be related only  to the   monthly distribution of the annual rainfall.  Comparison of models and observations    reveals a qualitative behavior  consistent with our interpretation of $D$, with  MRI-CGCM3 and CSIRO-Mk3-6-0
simulating  a too dry summer season and a too steep increase in December, resulting in a too short rainy season duration. On the other hand, MIROC5 and GISS-E2-R underestimate the rainfall fractions  during January-April and overestimate them during the dry period July-October,   resulting in a too flat $p_m$ annual distribution and negative  biases in RE.

 % +ve biases (SUB and WAM and EAS) 

\begin{figure*}[t]
\centering
  \noindent\includegraphics[trim = 0mm  80mm 20mm  0mm,clip,angle=90,width=0.9\textwidth,angle=-90]{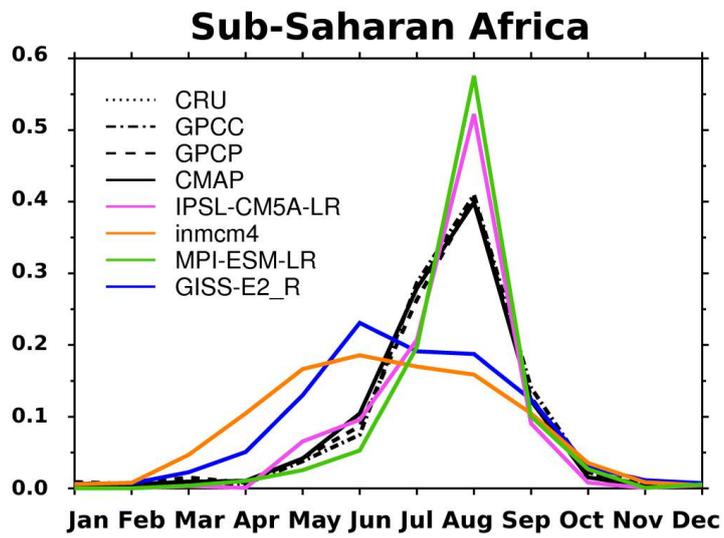}
  \caption{Precipitation fractions  for observation datasets CRU, GPCC, GPCP, CMAP and four models (IPSL-CM5A-LR, inmcm4, MPI-ESM-LR and GISS-E2-R) over Sub-Saharan Africa}\label{SUB}
\end{figure*}

\begin{figure*}[t]
\centering
  \noindent\includegraphics[trim = 0mm  0mm 0mm  0mm,clip,angle=90,width=0.9\textwidth,angle=-0]{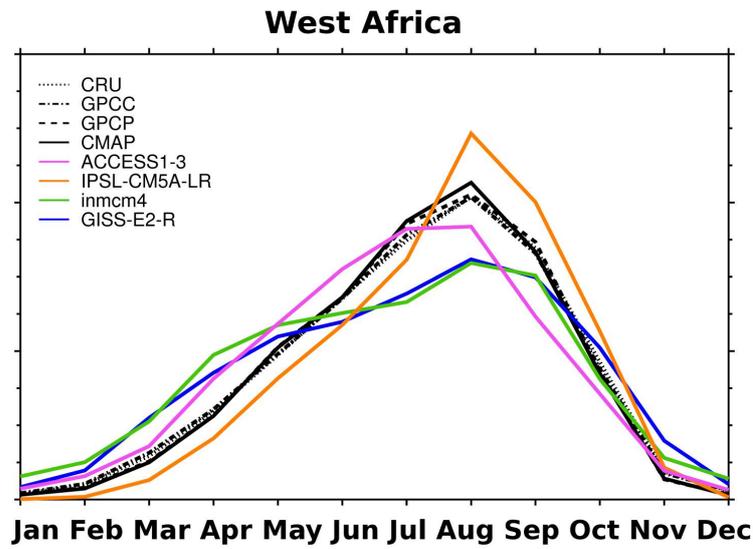}
  \caption{Precipitation fractions  for observation datasets CRU, GPCC, GPCP, CMAP and four models (IPSL-CM5A-LR, inmcm4, ACCESS1-3 and GISS-E2-R) over western Africa }\label{WAF}
\end{figure*}

Over the semi-arid Sub-Saharan region (Fig.~\ref{SUB}), observations show a strong   rainfall peak in August ($p_{\mathrm{aug}}\approx 0.4$) related to the marginal influence of the  West African monsoon  \citep{WAF} in the southern part of the region. As shown in Fig.~\ref{seconda}, these areas feature the highest RE values ($D\approx 1.6$) in the world. CMIP5 models generally underestimate RE in this region, except a few ones such as the two IPSL-CM5A and MPI-ESM models, which instead have positive biases in northern Africa. A direct inspection of their rainfall fractions reveal  that MPI-ESM-LR and IPSL-CM5A-LR simulate a  too pronounced  precipitation peak in August ($p_{\mathrm{aug}}\approx 0.5$ and $0.55$ respectively). Positive biases are instead associated with an overestimation of $p_m$ in late spring and an underestimation in summer. This tendency, which is common to most of CMIP5 GCMs, is evident from the   inmcm4 and GISS-E2-R (Fig.~\ref{SUB}), which are some of the models with the most severe negative biases ($\approx -0.5$).
In the  Western African monsoon region, south of the semi-arid Sub-Saharan Africa, negative biases in RE are less severe (Fig.~\ref{entr_all}). A few models have modest positive RE biases (IPSL-CM5A-LR, IPSL-CM5A-MR, MRI-CGCM3). A focus on this area (Fig.~\ref{WAF})  again elucidates the link between RE biases in terms of $p_m$.  The GISS-E2-R and the inmcm4 models, for example, overestimates $p_m$ in the dry season (November-April) and underestimates it during the wet season, resulting in a probability distribution $p_m$ more uniform, over the year, than  what is observed.  As an example we also show the rainfall fractions for a GCM (ACCESS1-3) which instead has almost no RE bias in this region. As expected,   the annual $p_m$ shape agrees relatively well with observations, apart from a slight shift of rainfall towards the early summer. The IPSL-CM5A-LR model, which tends to overestimate the RE, simulate a  too pronounced  precipitation peak in August. 
%Such a bias  thus points out to major models deficiencies to simulate the     large scale weather systems (e.g. the monsoons) delivering rainfall to  the region. 

\red{   %The GISS-E2-R model, in particular, overestimates $p_m$ in the dry season (November-April) and underestimates it during the wet season, resulting in a probability distribution $p_m$ more uniform, over the year, than  what is observed.  The   GISS-E2-R model, in particular,   has  one of  the largest RE negative bias over the  Western African monsoon region.  Similarly,  MIROC5 also overestimates $p_m$ in the dry season  and underestimates it during the wet season. A   negative RE  bias    therefore in this case  denotes a more spread out  annual rain distribution generally   associated with a positive precipitation  bias before the start and/or  after the end of the wet season.  Such a bias  thus points out to major models deficiencies to simulate the     large scale weather systems (e.g. the monsoons) delivering rainfall to  the region. 
}

\begin{figure*}[t]
\centering
  \noindent\includegraphics[trim = 0mm  80mm 0mm  30mm,clip,angle=0,width=0.8\textwidth,angle=-0]{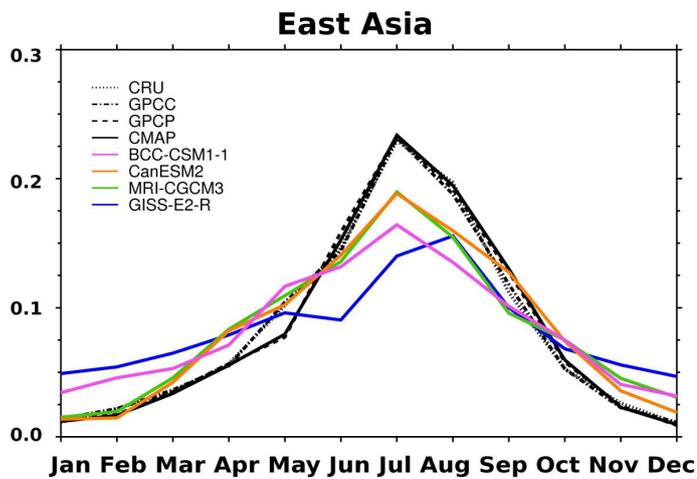}
  \caption{Precipitation frequencies $p_m=r_m/R$ for observation datasets CRU, GPCC, GPCP, CMAP and four models (BCC-CSM1-1, CanESM2, MRI-CGCM3 and GISS-E2-R)  over East Asia.  }\label{EAS}
\end{figure*}

 A persistent, large  negative RE bias ($\approx -0.5$), common to all models,  is also visible over  East Asia in most of the models shown in Fig.~\ref{entr_all}.
Direct inspection of $p_m$ shows that models do not capture the right magnitude of the July  peak in  $p_m$ and tend to have a too high rainfall fraction during the dry winter months. This behavior, which remains also in the least biased models (CanESM2 and MRI-CGM3),  is particularly evident in models with very large biases such as, for example,  GISS-E2-R and   BCC-CSM1-1 (Fig.~\ref{EAS}). The GISS model, in particular, considerably overestimate the winter rainfall fractions, resulting in a large bias in RE.

%Information provided by the RE about the duration of the wet season is in agreement with the estimations of the  rainy season duration provided by \cite{Kitoh} for the North African (longer),  South Asian (shorter), North American (agreement )  and East Asian monsoons (longer).

The RE can therefore  be a very useful metric  to test the right shape of the simulated monthly rain frequencies $p_m$ and to provide an estimation of the number of wet months.  It must be noted however that, from its  definition in Eq.~\ref{again},  the RE  is invariant to time translation ($p_m\rightarrow p_{m+s}$, with $s=1, \ldots 11$)  and therefore  not able to provide  information about  the onset and the decay  of the monsoon \citep{Sperber, Shabeh2}, which are   other two fundamental aspects of monsoonal regimes. Furthermore, the RE cannot  discriminate between regions with two short rainy season  and those with a single long one, since, from its own definition, any  reshuffle of the $p_m$ would not lead to any change in RE.

\subsection{Pattern correlation and Taylor diagrams}

 We conclude our comparison between CMIP5 models and observation by  estimating the pattern correlation (PC) and the \emph{centered} root mean square error (RMSE) between the simulated $R$, $D$, $S$ and the observed ones. The PC and the RMSE   are statistics generally  used to quantity pattern similarity    between two climatic fields ($f$, $r$) defined at $N$ points.  They  are defined as \citep[][]{TaylorDiag} $PC=[\sum (f_n-\overline{f})(r_n-\overline{r})]/(N\sigma_f \sigma_r)$ and $RMSE=\{\sum[(f_n-\overline{f})-(r_n-\overline{r})]^2/N \}^{1/2}$, where ($\overline{f}$, $\overline{r}$) and $\sigma_{f,r}$ are the  mean values and standard deviations of $f$ and $r$ respectively and  are related through the following relationship: $RMSE^2=\sigma_f^2+\sigma_r^2-2\sigma_f\sigma_r PC $. It has to be noted that since  the means are subtracted, the PC and the RMSE  cannot inform about overall biases (which have been analyzes in the previous sections instead) but just on the \emph{centered} pattern error. Since the GPCP and CMAP datasets are not bias-adjusted over oceans,  we restrict this comparison over land.   The GPCC land dataset is taken as  a reference  and compared to   CMIP5 models.  
 
\begin{figure*}
 \centering
\subfigure{\includegraphics[trim =0mm  10mm 10mm  70mm,clip,angle=0,width=0.9 \textwidth]{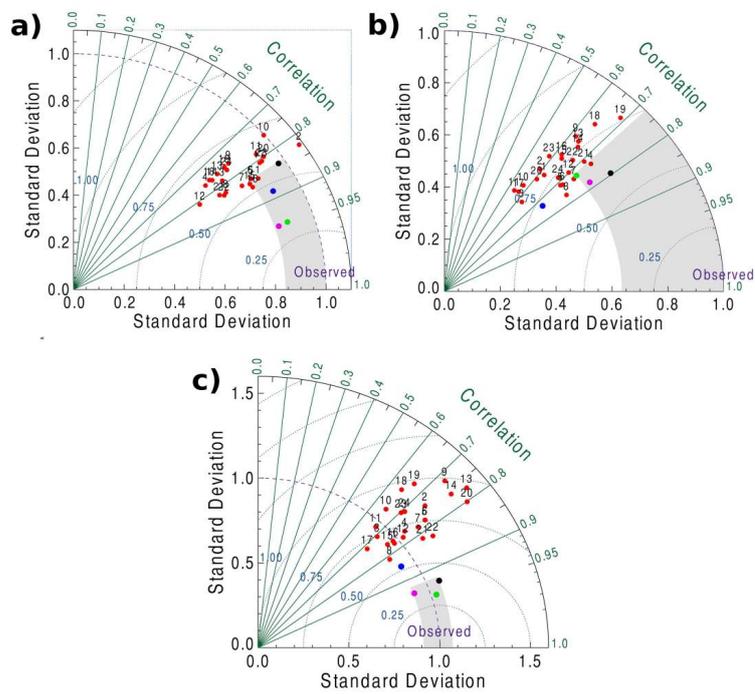} \label{tay1} }
\caption{Taylor diagrams for mean annual precipitation (a), relative entropy (b) and seasonality index (c). CMIP5 models are numbered as in Table~\ref{tab1}. The MME, CRU, GPCP and   CMAP are represented by the blues, black, green and magenta dots respectively. GPCC is taken as reference and standard deviations are normalized with respect to GPCC. The grey shaded areas are indicative of the range of observational uncertainty. The correlation analysis is restricted to land.  \label{taylor}}
\end{figure*}

 The values of the RMSE and PC for each of the CMIP5 models of Table~\ref{tab1} and the other precipitation gridded datasets are reported  in Table~\ref{tab3} and shown through   Taylor diagrams \citep{TaylorDiag} in Fig.~\ref{taylor}.  
   Other observational datasets  (CRU,  GPCP, CMAP) are also compared to GPCC and shown on the same diagrams in order to have   an indication about  observational uncertainty.   In fact, given the  problems in accurately measuring a highly spatially and temporally variable  field such as precipitation, observational  estimates are generally affected by  uncertainty   and more    observational  datasets   are needed to   provide information about the range of such  uncertainty.       To check   if the differences in model performances shown in Figure~\ref{taylor} are significant,       we considered, for a few models, all the five ensemble members available on the CMIP platform and  obtained  by initiating the simulations from different initial conditions. It is found that  the ensemble spread is very small and comparable with the size of the dots. 

In particular GPCP and CMAP are the closest to GPCC;  this is not surprising since these two satellite-based datasets use the GPCC dataset as their rain gauge component over land.
We define the  range  of observational uncertainty in terms of PC such as $[PC_{\mathrm{low}},1]$ and in term of  $\sigma$  such as  $[\sigma_{\mathrm{low}}, \sigma_{\mathrm{high}}]$  where $PC_{\mathrm{low}}$ is the lowest PC among the other observational datasets, $\sigma_{\mathrm{low}}=\inf \{\sigma_{\mathrm{gpcc}},\sigma_{\mathrm{obs}} \}$   and $\sigma_{\mathrm{high}}=\sup \{\sigma_{\mathrm{gpcc}},\sigma_{\mathrm{obs}} \}$. We have that  $PC_{\mathrm{low}}=0.83$, $[\sigma_{\mathrm{low}}  \sigma_{\mathrm{high}}]=[0.84,1]\sigma_{\mathrm{gpcc}}$ for mean annual precipitation;  $PC_{\mathrm{low}}=0.72$,  $[\sigma_{\mathrm{low}}, \sigma_{\mathrm{high}}]=[0.63,1]\sigma_{\mathrm{gpcc}}$ for the RE;   $PC_{\mathrm{low}}=0.92$, $[\sigma_{\mathrm{low}}, \sigma_{\mathrm{high}}]=[0.91,1.07]\sigma_{\mathrm{gpcc}}$ for the DSI (Fig.~\ref{tay1} and Table~\ref{tab3}).   A model therefore  performs   consistently  with observations  if its  $PC>PC_{\mathrm{low}}$ and its  $\sigma$  lies within the  range $[\sigma_{\mathrm{low}}, \sigma_{\mathrm{high}}]$.  Models  that perform worst are those with  $PC\ll PC_{\mathrm{low}}$, a  standard deviation $\sigma$ outside the range  $[\sigma_{\mathrm{low}}, \sigma_{\mathrm{high}}]$.   

In terms of precipitation, we note that most of the models   are placed outside the observational uncertainty range except ACCESS1-0, CCSM4 and CESM1-BGC  (Table~\ref{tab1}). While the MME does not necessarily have to outperform every single model \citep[e.g.][]{Sperber},   here this is the case  ($PC=0.88$, $\sigma=0.89\sigma_{\mathrm{gpcc}}$) and it is consistent with the observations. HadGEM2-ES, HadGEM2-CC and CESM1-CAM5 also perform well with a PC larger or equal  than $0.83$ but with a value of the    standard deviation  just below $\sigma_{\mathrm{low}}=0.84\sigma_{\mathrm{gpcc}}$. Overall,  other GCMs are placed not far from the lower bound of the PC ($\approx 0.8$), but some of them underestimate $\sigma$ by a factor $0.3$ or more  (e.g. GFDL-CM3, GFDL-ESM2M, IPSL-CM5A-LR), well below $\sigma_{\mathrm{low}}=0.83\sigma_{\mathrm{gpcc}}$, resulting in large RMSE (CanESM2, GFDL-ESM2G, IPSL-CM5-LR). 
As far as RE is concerned, observational  uncertainty is generally  larger ($PC_{\mathrm{low}}=0.72$, $\sigma_{\mathrm{low}}=0.63\sigma_{\mathrm{gpcc}}$). This is  consistent with the  large differences between the CRU and GPCC datasets shown in Fig.~\ref{terza}.  CanESM2 ($PC=0.73$, $\sigma=0.71$) and CESM1-CAM5 ($PC=0.73$, $\sigma=0.63$) are within the observational range  range whereas  CNRMS-CM5 is  just slightly outside ($PC=0.57$). Contrary to the case of precipitation, the MME for the RE  lies outside such range and does not outperform every single model. Some of the CMIP5 models perform particularly badly  and feature a considerably lower PC and $\sigma$, resulting in RMSE almost comparable with $\sigma$ (GISS-E2-H, GISS-E2-R, MRI-CGM3).
Again,   most of the  model  are  not far from the  lower bounds of  observational uncertainty. It is interesting to note that the best performing models in terms of the field of mean annual precipitation are not the best in terms of RE.
Finally we note that  no model is consistent with observations in terms of the DSI, as evident from Fig.~\ref{taylor}.  This may be due to the fact that the DSI is a diagnostic metrics more complex than precipitation and RE  alone --  it combines them together, providing integrated information about  the intensity of the annual rainfall and the shape of the monthly rain frequency -- and therefore it is more unlikely for models to capture equally well spatial variability of both precipitation and RE.  Rainfall is a complex field and it is challenging  for models to properly simulate it. Lack of  model agreement  between mean  precipitation and other, more complex aspects  are found also, for example,  when comparing total precipitation and upper quantiles of the precipitation distribution. For example,    analyzing  CMIP5 models, \cite{Mehran} show  that   models best simulating the total precipitation amounts not necessarily   are also  the  best performing in precipitation upper quantiles.  

   The MME outperforms every single model but still lies outside the observational range. 
When all three metrics are considered, CESM1-CAM5 is overall one of best   model    in terms of spatial variability  and magnitude (as evident from Fig.~\ref{entr_all}--Fig.~\ref{intercomparison}) whereas the worst performing models are  GISS-E2-H and GISS-E2-R.

\section{Conclusions}
\label{finale}

 Future  improvements and developments of the GCM representation of precipitations  strongly rely on
 rigorous metrics for their validation \citep[e.g. ][]{Mehran}.  Accurate, reliable diagnostics of  rainfall seasonality  is a 
 necessary tool for gauging  GCMs performance, evaluating  their  realism  and  quantifying  changes in the hydroclimatic regimes.    In this study we  used  novel  measures of rainfall seasonality  \citep{Porporato} based on information entropy, namely the relative entropy (RE) and the dimensionless seasonality index (DSI),   for characterizing  the seasonality of precipitation regimes  during the 1950-2010 period over lands and oceans using the  four  recently updated precipitation gridded datasets   GPCC, CRU, CMAP, GPCP  (Fig.~\ref{seconda} and Fig.~\ref{sesta})  and for  assessing    CMIP5 models' ability  capture the observed patterns of RE and DSI.
   
The RE  provides an integral  measure  of     the seasonality of  the annual rainfall curve 
whereas the DSI  quantifies  the intensity of the rainfall during the wet  season.  
The RE is related  to  the number   of the wet time accumulation bins $n^\prime$ through the   simple relation $n^\prime \approx N\cdot 2^{-D}$, where  $N$  is the number of temporal  accumulation bins in a year. Areas with high RE are therefore characterized by   prolonged dry periods and rainfall concentrated in a short time period. Given its own definition, the RE cannot discriminate between unimodal and bimodal rainfall regimes and therefore it does not automatically provide a measure of the duration of the wet season. However,  for precipitation regimes known to be unimodal (e.g. in the South Asian monsoon region), $n^\prime$ coincides with  the duration of the wet season and it can be used as a further measure along with more tradition ones such as the monsoon retreat and onset time.

It is found that Equatorial  (Indonesia, Congo basin, Amazon) and midlatitude regions  have low values of  the DSI because, in spite of the large mean annual precipitation,  their relative entropy is very small ($\le 0.05$). Arid and semiarid regions around $20^\circ$ N with intermittent precipitation regimes -- like the sub-Saharan Sahel -- are characterized by large  RE   and feature very low DSI  because of the very little annual precipitation. Highest DSI ($\ge 0.05$) are therefore found in those regions with intermediate-to-high  levels of mean annual rainfall and RE such as 
 northeast region of Brazil, Western Africa, Northern Australia,   Western Mexico and  South-Southeast and Eastern Asia, which constitute the so-called global monsoon region \citep{Ding}.
According to the DSI,    the west coast of North America, Mediterranean, Middle-East regions and  the  Andes also  feature appreciable seasonality,  since rainfall in these regions is  confined to the (local) winter months.

The RE and  DSI have two practical advantageous features: a) the robustness against changes 
 of  the accumulation temporal bin  of  the precipitation time  series and b) the  coarse-graining properties of the  RE (Fig.~\ref{pent}). 
 The first property  guarantees   a  quantification  of seasonality which is as much as possible  independent of   the time bin (day, pentad, week, month)
 used to accumulate precipitation.   The second property allows us   to establish  lower bounds for the RE and DSI with respect to  values 
 which would be   obtained    from  higher-resolution  data, which are not always available.    
      
Comparison of simulations performed with 24    CMIP5 coupled atmosphere-ocean general circulation models with  
the   precipitation datasets   over the period 1950-2010 reveals consistent positive (South America) and negative (East Asia, northern Africa) RE biases across models (Fig.~\ref{entr_bias}). Such biases are related to GCMs'  inability  to simulate the right monthly fractions of rainfall along the year.  The GCMs'  negative RE bias  over western Africa is due to  a positive
 precipitation bias in the West African monsoon region in late spring and a negative precipitation bias during July-September (Fig.~\ref{SUB} and ~\ref{WAF}).  A similarly  consistent picture has been shown to explain also  the large   negative bias in east Asia, related to rainfall fractions which are too low during the wet May-September period and too large during the dry October-April period, thus resulting in a $p_m$ sequence not peaked enough (Fig.~\ref{EAS}).  On the other hand, the positive RE  bias  over tropical southern American (Fig.~\ref{SAM})  is due to the opposite tendency to overestimate the monthly rainfall fractions during the local summer and underestimate them in the late spring/early summer period, resulting in an excessively peaked $p_m$.
  The presence  of these RE biases consistently across the evaluated CMIP5 GCMs  indicate  the presence of general deficiencies in the models in simulating tropical precipitation and, in particular, monsoons \citep{Turner}.  These systematic RE  errors  appear  to be not very sensitive to differences in model horizontal resolution since they are found in models with higher and lower space resolution and are likely to be   due general shortcomings in representing the dynamics or physics of climatic phenomena.

In terms of spatial variability, pattern correlation analysis over continents clearly shows that CMIP5 models 
have a better skill in reproducing the variability pattern of precipitation compared to RE  (Fig.~\ref{taylor}) with few models   consistent with observations. In particular,  no model reproduces the DSI spatial variability consistently with observations.  Overall, CESM1-CAM5 is  one of the best performing  models  for all three metrics, whereas the worst performing are  GISS-E2-H and GISS-E2-R.

It has to be noted  that  
 RE and DSI  do not provide a complete description of rainfall seasonality since they do not take into account the timing of the wet season.  Their main scope  is to provide an  easy way to compare maps of RE/DSI between various datasets to  determine areas of interest, and then to undertake a more detailed analysis of seasonality in these regions using more traditional measures of
seasonality as we have shown for  the West African, the Australian,  the southern American and the eastern Asian regions.

The  methodology underlying the definition of the RE and DSI is 
 very general  and applicable to much more general  cases than what shown in this study. In principle it may be   adapted to other  periodic or quasi-periodic climatic  sequences of  positive-definite variables. Such tools  seems therefore very promising for assessing models'  capability to simulate spatial and temporal   patterns of  the rainfall diurnal cycle, which is the primary mode of variability  in the equatorial regions, where heavy rains are concentrated in the afternoon hours.  Furthermore, the diagnostic tools presented in this
   study can  be used for studying changes  in rainfall seasonality for future climate projections under 
   anthropogenic forcing in addition to  more traditional approaches \citep{Huang, Wang, LeeWang}.      Future efforts in this direction  will focus therefore on the application to the rainfall diurnal cycle  and on the  analysis of   21st century greenhouse-forced  climate projections.

\clearpage

%
%\begin{appendix}% Use \begin{appendix} and not \begin{appendix}[A] for only one appendix.
%\label{appendixA}

\appendix 

\section{ Properties of the relative entropy }
\label{appendix}

\subsection{Relative entropy and  information entropy} 
\label{A1}

Given a discrete probability distribution $p=\{p_m\}_{m=1}^N$ describing a random variable, the \emph{information entropy} associated with  ${p_m}$   is a measure of the uncertainty of a random variable described by $p$ and  it   is defined as 
\begin{equation}
\mathcal{H}\left(p\right)\equiv -\sum_m^N p_m\log_2 p_m
\label{app1}
\end{equation}
and   $0\le \mathcal{H} \le \max(\mathcal{H})$, where  $\max(\mathcal{H})=\log_2 N$ for the uniform distribution   $p_m=1/N$  (\emph{maximum uncertainty})  and 0 if one out of the $N$ values of $p$ is equal to one  and all the remaining are zero ($x\log x\rightarrow 0$ as $x\rightarrow 0$)(\emph{no uncertainty}). In the case considered in this study, $N=12$ and $\max(\mathcal{H})=\log_2 12$. The \emph{relative entropy}  of $p$ with respect to $q$, $\mathcal{D}(p|q)$,  is introduced instead to measure how different  two probability distribution $\{p_m\}_{m=1}^N$ and $\{q_m\}_{m=1}^N$ are:
\begin{equation}
\mathcal{D}(p|q)\equiv \sum_{m=1}^N p_m\log_2\left(\frac{p_m}{q_m}\right)
\label{app2}
\end{equation}
and it measures the inefficiency of assuming $q$ when instead the true distribution is $p$.   It can be demonstrated \citep{Cover} that $\mathcal{D}(p|q)\ge 0$ for any $p, q$ and $\mathcal{D}(p|q)=0$ if and only if the two probability distributions are the same. The relative entropy is not symmetric, $\mathcal{D}(p|q) \ne \mathcal{D}(q|p)$ and therefore is not a distance in  a mathematical sense. However  it is still useful to think of it as a distance between probability distributions. For defining the seasonality index we define $D\left(p\right)$ such as
\begin{equation}
D(p)\equiv \sum_{m=1}^{N} p_m\log_2\left( N\,p_m \right)
\label{app2_bis}
\end{equation}
that is such as the relative entropy of   the probability distribution $p_m$  with respect to   the  uniform distribution $q_m=1/N$, which is taken as a reference.  In the following and in the rest of this manuscript  we will still refer to $D(p)$ as relative entropy.     From this definition it follows that
\begin{equation}
D\left(p\right)=-\mathcal{H}\left(p\right)+\log_2 N.
\label{app3}
\end{equation}
As a consequence, for two probability distributions $p$ and $w$, $D\left(p\right)-D(w)=\mathcal{H}(w)-\mathcal{H}\left(p \right)$. 

\subsection{Relative entropy and the spread of $p_m$} 
\label{A2}

Let us assume now that $\{p_m\}_1^{N}$ are the monthly precipitation fractions ($N=12$). From what said so far,   it is expected that the larger it is  $D$, the less   uniformly    the precipitation is distributed throughout the year. So $D$ is related to the ``spread'' of precipitation signal. This concept can be framed in a rigorous way in information theory  by defining the  \emph{effective} number of values of $p$
\begin{equation}
n^\prime\left(p\right)=2^{\mathcal{H}\left(p\right)}=12\, \cdot 2^{-D}.
\label{app4}
\end{equation}
Mathematically $n^\prime$ defines the   number of months over which $p_m$ is considerably different from zero, i.e. the support of $p_m$ \citep{Cover}. Therefore  $n^\prime$  can be interpreted as the \emph{effective} number of wet months in a year. Areas  characterized by  $D=0$  have $n^\prime = 12$, that is  no significant dry period (non-seasonal rainfall regime), whereas regions featuring $D=D_{\mathrm{max}}=\log_2 12$, $p_k=1$ have their annual precipitation all concentrated in one month (extreme seasonal rainfall regime).
%From (\ref{app4}) the \emph{entropic spread} $E$ is defined as $E=[\left(n^{\prime 2}-1\right)/12]^{1/2}$ (that is as the standard deviation of a uniform distribution with $n^\prime$ values) and  it can be readily shown (eqs. \ref{app3} and \ref{app4}) that
%\begin{equation}
%\label{app5}
%E =  [12\,\cdot 2^{-2D}-1/12]^{1/2}.        
%\end{equation}
%Consequently $E$ varies between $0$ (single-month probability distribution)  and $\sqrt{12-1/2}\approx 3.4$ (uniform rain  distribution). 
 For regions having a unimodal seasonal rainfall distribution,  $n^\prime$ provides   a measure of the duration of the wet season (e.g. Indian region). It has to be noted however that different measures of the wet season duration which are  not based on integral properties of the rainfall distribution  but on local properties -- e.g. retreat minus onset dates \citep{Sperber, Kitoh, Shabeh2}, where onset and retreat are defined by the $5$ mm day$^{-1}$ threshold -- may give different results.

\begin{figure*}
 \centering
\subfigure{\includegraphics[angle=0, width=0.8\textwidth]{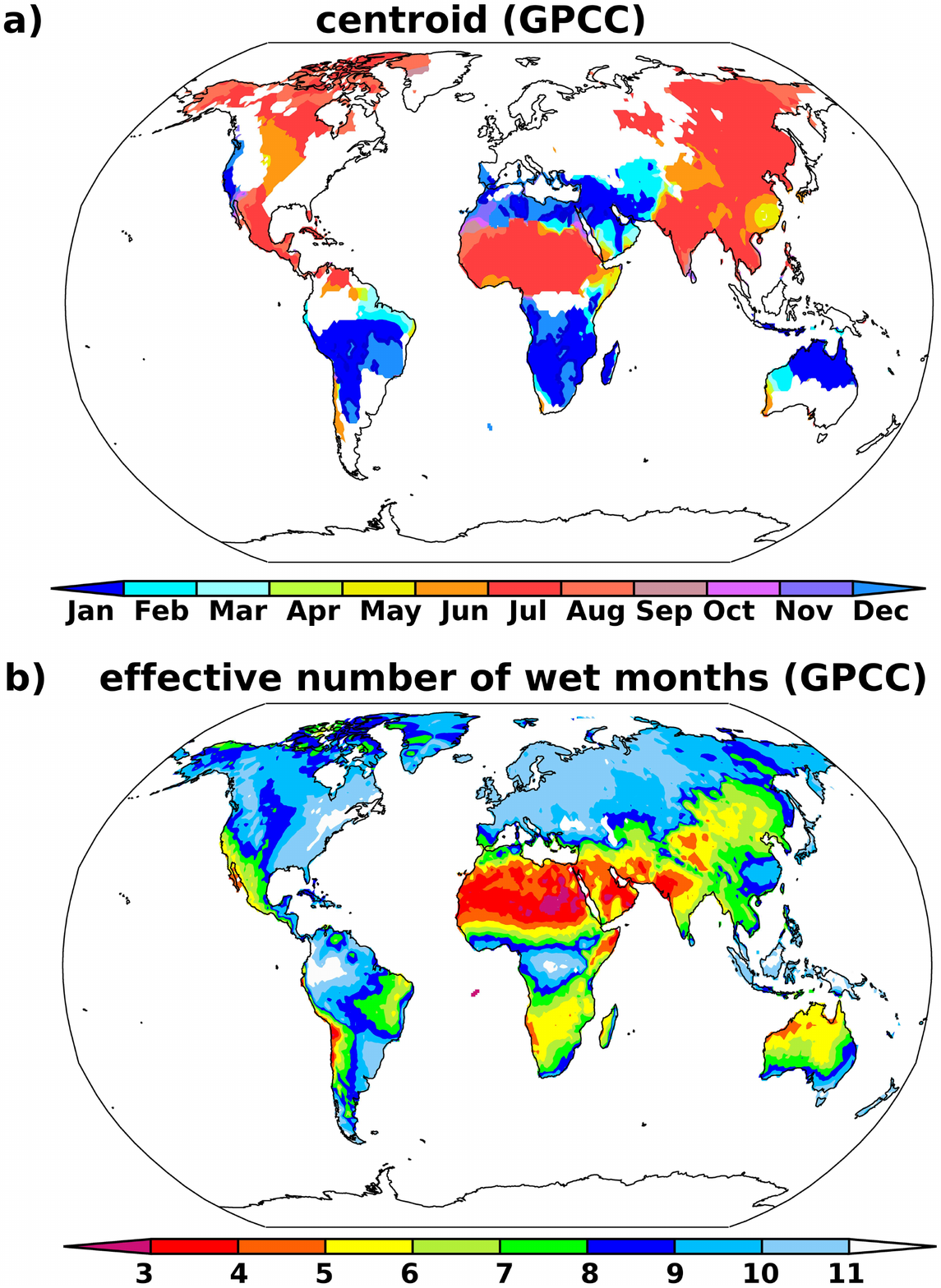} }
\caption{Centroid (a) and effective number of wet months  (b)  of the monthly precipitation sequence $p_m$  for the GPCC dataset. In (a) numbers denotes the months of the year around which $p_m$ is centered and in (b) the length in months.  The centroid is not shown for regions with $D\ge 0.2$ -- corresponding approximately to $n^\prime$ greater than $11$. \label{circular}}
\end{figure*}

Within this framework, let us also introduce another useful statistical indicator of rainfall seasonality, the \emph{centroid}.  By using circular statistics \citep{circular},  the   first moment of $p_m$ (centroid) is defined as 
\begin{equation}
C=\textrm{arg}(z), \quad z=\sum_{m=1}^{12} p_m e^{i \frac{2 \pi m}{12} }
\end{equation}
and it is shown in Fig.~\ref{circular}. The centroid provides a measure of the the timing of the wet season. While it can be mathematically defined for any precipitation sequence $r_m$ and so in any location, it is really meaningful only for those rainfall regimes that are somewhat ``localized'' during the year -- i.e. having a clear dry and wet period. A more extensive analysis of $C$ in present condition and future emission scenarios will be reported elsewhere. 
%and the circular standard deviation as
%\begin{equation}
%\sigma=\sqrt{\ln(1/|z|^2)}
%\label{spreadcirc}
%\end{equation}
%The precipitation centroid $C$, the spread $\sigma$ and the entropic spread $E$ are shown in Fig.~\ref{quinta}. 
%and are shown in Fig.~\ref{circular}. 

%For regions having a unimodal seasonal rainfall distribution,  $\sigma$ provides   a measure of the duration of the wet season $l$, the following relation is found to approximately hold
%\begin{equation}
%l\approx n^\prime \approx 2\sigma.
%\label{dur}
%\end{equation}
  %     It has to be noted however that different measures of the wet season duration which are  not based on integral properties of the rainfall distribution  but on local properties -- e.g. retreat minus onset dates \citep{Sperber, Kitoh, Shabeh2}, where onset and retreat are defined by the $5$ mm day$^{-1}$ threshold -- may give different results

\subsection{Coarse graining properties of $D$}
\label{A3}

\begin{figure*}
\centering
  \subfigure{\includegraphics[trim = 0mm  0mm 0mm  0mm,clip,angle=0,width=0.7\textwidth,angle=0]{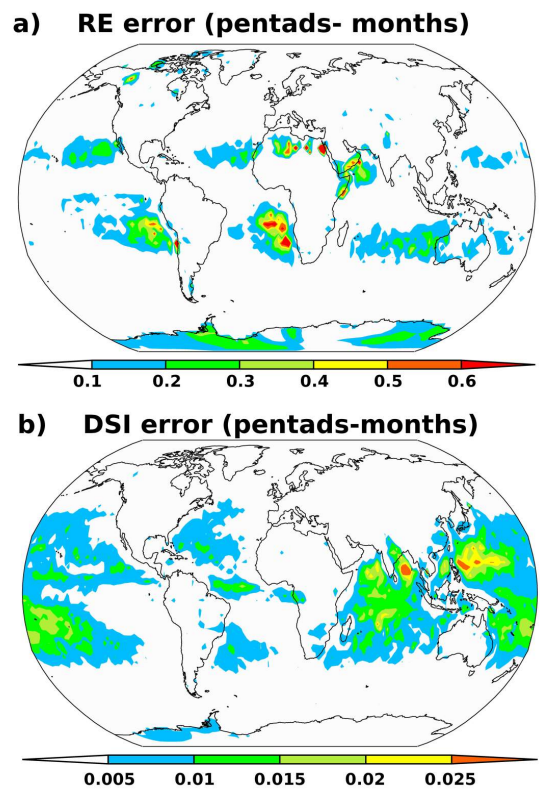} }
    \caption{ Difference between the RE (a)  and the DSI (b)  estimated from pentad and monthly means. }\label{pent}
\end{figure*}

A remarkable property of $D$ is the possibility to control its magnitude as the time resolution of the time series is coarse grained.
  The choice of a certain time series resolution is somewhat 
 arbitrary and dependent on the data available. It is therefore  desirable to have indicators that are stable against changes in the
 accumulation time bin or, at least, that vary in a controllable way. Relative entropy allows us to set lower bounds for the
 error associated with  the loss of information due to time coarse-graining. If $D_N$ is the relative
 entropy estimated from rainfall data at high time resolution $\tilde{p}_j$  (e.g. daily, $N = 365$ or
  pentads, $N=73$), we can
 aggregate sequentially $\nu$ of the $\tilde{p}_j$  (e.g. $\nu=5$   for pentads) and obtain

\begin{equation}
p_i=\sum_{j=\nu i-\nu+1}^{\nu i} \tilde{p}_j 
\end{equation}
with $i=1,\ldots M$ and $M=N/\nu$. By using  the \emph{log sum inequality} \citep{Cover}
\begin{equation}
\sum_{i=1}^{n} a_i\log{\frac{a_i}{b_i}}\ge \left( \sum_{i=1}^{n} a_i \right)\log{\frac{\sum_{i=1}^n a_i}{\sum_{i=1}^n b_i}}, \quad a_i, b_i\ge 0
\label{logsum}
\end{equation}  
where the equality holds only if the $a_i$ and the $b_i$ do not depend on $i$,   and from  the definition (\ref{app2_bis}) it follows that
\begin{equation}
\sum_{j=1}^{N} \tilde{p}_j \log_2{\frac{\tilde{p}_j}{\tilde{q}_j}} \ge \sum_{i=1}^{N/\nu} \left( \sum_{j=\nu i-\nu+1 }^{\nu i } \tilde{p}_j \right) \left( \log_2 \frac{\sum_{j=\nu i-\nu+1}^{\nu i} \tilde{p}_j}{\sum_{j=\nu i-\nu+1}^{\nu i} \tilde{q}_j} \right)=\sum_{i=1}^{M} p_i\log_2 (p_i/q_i)
\label{logsum2}
\end{equation}
and therefore
\begin{equation}
D_{N} \ge D_{M} \quad  \textrm{for} \quad N\ge M.
\label{dis}
\end{equation}
From the definition of the dimensionless seasonality index $S$ in Sect.~\ref{data_methods} (Equation~\ref{dsi}), it
 is obvious that also $S_N \ge S_M$ and so information about rainfall seasonality is lost in the
 upscaling procedure unless the values in each temporal bin are equal. In Fig.~\ref{pent} the differences $D_{73}-D_{12}$ and 
$S_{73}-S_{12}$  are shown as an example. 

%\end{appendix}

%========================================================================
%                                                             FIGURES
%========================================================================

\begin{acknowledgements}

The authors  acknowledge the World Climate Research ProgrammeÕs Working Group on Coupled Modeling, which is responsible for CMIP,  and 
 the NOAA/OAR/ESRL PSD, Boulder, Colorado, USA,  for providing from their Web site  the  CMAP, GPCP  and GPCC precipitation data.   SP, VL  and SH wish to acknowledge the financial support provided by the ERC-Starting Investigator Grant NAMASTE (Grant no. 257106) and by the CliSAP/Cluster of excellence in  the Integrated Climate System Analysis and Prediction.  AP gratefully acknowledges NSF Grants: 
CBET 1033467,  EAR 1331846,  EAR 1316258 as well as the US DOE through the Office of Biological and Environmental Research, Terrestrial Carbon Processes program (DE-SC0006967), the Agriculture and Food Research Initiative from the USDA National Institute of Food and Agriculture (2011-67003-30222). XF  acknowledges funding from the NSF Graduate Research Fellowship Program. F. Ragone, J. M. Gregory, G. Badin and F. Lalibert\'e  are thanked for useful comments and suggestions.  The authors also wish  to thank B. G. Liepert and F. Lo 
for providing numerical data about CMIP5 models water biases and two anonymous reviewers for their constructive suggestions which helped us to improve this manuscript.

\end{acknowledgements}

\clearpage

%\bibliography{references.bib}   % name your BibTeX data base

\begin{thebibliography}{75}
\providecommand{\natexlab}[1]{#1}
\providecommand{\url}[1]{{#1}}
\providecommand{\urlprefix}{URL }
\expandafter\ifx\csname urlstyle\endcsname\relax
  \providecommand{\doi}[1]{DOI~\discretionary{}{}{}#1}\else
  \providecommand{\doi}{DOI~\discretionary{}{}{}\begingroup
  \urlstyle{rm}\Url}\fi
\providecommand{\eprint}[2][]{\url{#2}}

\bibitem[{Allen and Ingram(2002)}]{Allen}
Allen MR, Ingram WJ (2002) Constraints on future changes in climate and the
  hydrological cycle. Nature 419:224--232

\bibitem[{Becker et~al(2013)Becker, Finger, Meyer-Christoffer, Rudolf, Schamm,
  Schneider, and Ziese}]{GPCC2}
Becker E, Finger P, Meyer-Christoffer A, Rudolf B, Schamm K, Schneider U, Ziese
  M (2013) A description of the global land-surface precipitation data products
  of the global precipitation climatology centre with sample applications
  including centennial (trend) analysis from 1902-present. Earth Syst Sci Data
  71:71--99

\bibitem[{Bengtsson et~al(2006)Bengtsson, Hodges, and Roeckner}]{Bengtsson}
Bengtsson L, Hodges KI, Roeckner E (2006) Storm tracks and climate change.
  Journal of Climate 19:3518 -- 3543

\bibitem[{Bolvin et~al(2009)Bolvin, Huffman, Nelkin, and Poutiainen}]{Bolvin}
Bolvin DT, Huffman RFAGJ, Nelkin EJ, Poutiainen JP (2009) Comparison of {GPCP}
  monthly and daily precipitation estimates with high-latitude gauge
  observations. Journal of Applied Meteorology and Climatology 48(9):1843 --
  1857

\bibitem[{Boos and Hurley(2013)}]{Boos}
Boos WR, Hurley JV (2013) Thermodynamic bias in the multimodel mean boreal
  summer monsoon. J Climate 26:2279 -- 2287

\bibitem[{Borchert(1994)}]{Borchert}
Borchert R (1994) Soil and stem water storage determine phenology and
  distribution of tropical dry forest trees. Ecology 75:1437 -- 1449

\bibitem[{Camargo(2013)}]{Camargo}
Camargo S (2013) Global and regional aspects of tropical cyclone activity in
  the cmip5 models. J Climate 26:9880 -- 9902

\bibitem[{Chadwick et~al(2013)Chadwick, Boutle, and Martin}]{Chadwick}
Chadwick R, Boutle I, Martin G (2013) Spatial patterns of precipitation change
  in {CMIP}5. J Climate 26:3803 -- 3822

\bibitem[{Cherchi et~al(2011)Cherchi, Alessandri, Masina, and
  Navarra}]{Cherchi}
Cherchi A, Alessandri A, Masina S, Navarra A (2011) Effects of increased
  {CO}$_2$ levels on monsoons. Climate Dynamics 37:83--101

\bibitem[{Chou et~al(2009)Chou, Neelin, Chen, and Tu}]{Chou2}
Chou C, Neelin JD, Chen CA, Tu JY (2009) Evaluating the ``rich-get-richer''
  mechanism in tropical precipitation change under global warming. J Climate
  22:1982 -- 2005

\bibitem[{Chou et~al(2013)Chou, Chiang, Lan, Chung, Liao, and Lee}]{Chou}
Chou C, Chiang JCH, Lan CW, Chung CH, Liao YC, Lee CJ (2013) Increase in the
  range between wet and dry season precipitation. Nature Geoscience 6:263 --
  267, \doi{10.1038/NGEO1744}

\bibitem[{Christensen et~al(2008)Christensen, Boberg, Christensen, and
  Lucas-Picher}]{Christensen}
Christensen JH, Boberg F, Christensen OB, Lucas-Picher P (2008) On the need for
  bias correction of regional climate change projections of temperature and
  precipitation. Geophys Res Lett 35:L20,709, \doi{10.1029/2008GL035694}

\bibitem[{Cook and Seager(2013)}]{Cook}
Cook BI, Seager R (2013) The response of the north american monsoon to
  increased greenhouse gas forcing. J Geophys Res 118:1690 -- 1699

\bibitem[{Cover and Thomas(1991)}]{Cover}
Cover TM, Thomas JA (1991) Elements of Information Theory. Wiley-Interscience

\bibitem[{Deidda et~al(1999)Deidda, Benzi, and Siccardi}]{Deidda}
Deidda R, Benzi R, Siccardi F (1999) Multifractal modeling of anomalous scaling
  laws in rainfall. Water Resources Research 35(6):1853 -- 1867

\bibitem[{Eamus(1999)}]{Eamus}
Eamus D (1999) Ecophysiological traits of deciduous and evergreen woody species
  in the seasonally dry tropics. Trends Ecol Evol 14:11 -- 16

\bibitem[{Feng et~al(2013)Feng, Porporato, and Rodriguz-Iturbe}]{Porporato}
Feng X, Porporato A, Rodriguz-Iturbe I (2013) Changes in rainfall seasonality
  in the tropics. Nature Climate Change \doi{10.1038/nclimate1907}

\bibitem[{Fisher et~al(1993)Fisher, Lewis, and Embleton}]{circular}
Fisher N, Lewis T, Embleton BJJ (1993) Statistical Analysis of Spherical Data.
  Cambridge University Press

\bibitem[{Frierson et~al(2013)Frierson, Hwang, Fuckar, Seager, Kang, Donohoe,
  Maroon, Liu, and Battisti}]{Frierson}
Frierson DMW, Hwang YT, Fuckar NS, Seager R, Kang SM, Donohoe A, Maroon EA, Liu
  X, Battisti DS (2013) Contribution of ocean overturning circulation to
  tropical rainfall peak in the northern hemisphere. Nature Geoscience 6:940 --
  944

\bibitem[{Goswami et~al(1999)Goswami, Krishnamurthy, and Annamalai}]{Goswami}
Goswami BN, Krishnamurthy V, Annamalai H (1999) A broad-scale circulation index
  for the interannual variability of the indian summer monsoon. Quart J Roy
  Meteor Soc 125:611--633

\bibitem[{Guilyardi et~al(2013)Guilyardi, Balaji, Lawrence, Callaghan, Deluca,
  Denvil, Lautenschlager, Morgan, Murphy, and Taylor}]{Gui}
Guilyardi E, Balaji V, Lawrence B, Callaghan S, Deluca C, Denvil S,
  Lautenschlager M, Morgan M, Murphy S, Taylor KE (2013) Documenting climate
  models and their simulations. Bull Amer Meteor Soc 94:623 -- 627

\bibitem[{Harris et~al(2013)Harris, Jones, Osborn, and Lister}]{CRU}
Harris I, Jones PD, Osborn TJ, Lister DH (2013) Updated high-resolution grids
  of monthly climatic observations. Int J Climatol p in press,
  \doi{10.1038/NGEO1744}

\bibitem[{Harvey et~al(2012)Harvey, Shaffrey, Woolings, Zappa, and
  Hodges}]{Harvey}
Harvey BJ, Shaffrey LC, Woolings TJ, Zappa G, Hodges KI (2012) How large are
  projected 21st century storm track changes? Geophysical Research Letters
  39:L18,707, \doi{10.1029/2012GL052873}

\bibitem[{Hasson et~al(2013)Hasson, Lucarini, and Pascale}]{Shabeh1}
Hasson S, Lucarini V, Pascale S (2013) Hydrologycal cycle over south and
  southeast asian river basins as simulated by {PCMDI/CMIP}3 experiments. Earth
  System Dynamics 4:199--217

\bibitem[{Hasson et~al(2014)Hasson, Lucarini, Pascale, and B\"ohner}]{Shabeh2}
Hasson S, Lucarini V, Pascale S, B\"ohner J (2014) Seasonality of the
  hydrologycal cycle in major south and southeast asian river basins as
  simulated by {PCMDI/CMIP}3 experiments. Earth System Dynamics p in press

\bibitem[{Held and Soden(2006)}]{Soden}
Held IM, Soden BJ (2006) Robust responses of the hydrological cycle to global
  warming. J Climate 19:5686--5699

\bibitem[{Huang et~al(2013)Huang, Xie, Hu, Huang, and Huang}]{Huang}
Huang P, Xie SP, Hu K, Huang G, Huang R (2013) Patterns of the seasonal
  response of tropical rainfall to global warming. Nature Geoscience 6:357 --
  361

\bibitem[{Huffman et~al(2009)Huffman, Adler, Bolvin, and Gu}]{Huffman}
Huffman GJ, Adler RF, Bolvin DT, Gu G (2009) Improving the global precipitation
  records: {GPCP} version 2.1. Geophys Res Lett 36(17):L17,808

\bibitem[{Hwang and Frierson(2013{\natexlab{a}})}]{Hwang}
Hwang YT, Frierson DMW (2013{\natexlab{a}}) Link between the
  double-intertropical convergence zone problem and cloud biases over the
  southern ocean. PNAS \doi{10.1073/pnas.1213302110}

\bibitem[{Hwang and Frierson(2013{\natexlab{b}})}]{HwangSupp}
Hwang YT, Frierson DMW (2013{\natexlab{b}}) Link between the
  double-intertropical convergence zone problem and cloud biases over the
  southern ocean: Supplementary material. PNAS
  \doi{www.pnas.org/lookup/suppl/doi:10.1073/pnas.1213302110/-/DCSupplemental.}

\bibitem[{IPCC(2013)}]{IPCC5}
IPCC (2013) IPCC Fifth Assessment Report: Working Group I Report "The Physical
  Science Basis". Cambridge University Press

\bibitem[{Kajikawa et~al(2010)Kajikawa, Wang, and Yang}]{Kajikawa}
Kajikawa Y, Wang B, Yang J (2010) A multi-time scale australian monsoon index.
  Int J Climatol 30:1144--1120

\bibitem[{Kang and Lu(2012)}]{Kang}
Kang SM, Lu J (2012) Expansion of the hadley cell under global warming: Winter
  versus summer. J Climate 25:8387Ð8393

\bibitem[{Kelley et~al(2012)Kelley, Ting, Seager, and Kushnir}]{Kelley}
Kelley C, Ting M, Seager R, Kushnir Y (2012) Mediterranen precipitation
  climatology, seasonal cycle, and trend as simulated by {CMIP}5. Geophysical
  Research Letters 39:L21,703, \doi{10.1029/2012GL053416}

\bibitem[{Kharin et~al(2013)Kharin, Zwiers, Zhang, and Wehner}]{Kharin}
Kharin VV, Zwiers FW, Zhang X, Wehner M (2013) Changes in temperature and
  precipitation extremes in the {CMIP}5 ensemble. Climatic Change 119(2):345 --
  357

\bibitem[{Kitoh et~al(2013)Kitoh, Endo, Kumar, and Cavalcanti}]{Kitoh}
Kitoh A, Endo H, Kumar KK, Cavalcanti IFA (2013) Monsoons in a changing world:
  a regional perspective in a global context. Journal of Geophysical Research
  118:1--13

\bibitem[{Knutti(2010)}]{Knutti}
Knutti R (2010) The end of model democracy? Climatic Change 102:395 -- 404

\bibitem[{Konar et~al(2010)Konar, Muneepeerakul, Azaele, Bertuzzo, Rinaldo, and
  Rodriguez-Iturbe}]{Itur}
Konar M, Muneepeerakul R, Azaele S, Bertuzzo E, Rinaldo A, Rodriguez-Iturbe I
  (2010) Potential impacts of precipitation change on large-scale patterns of
  tree diversity. Water Resour Res 46:W11,515, \doi{10.1029/2010WR009384}

\bibitem[{Lee and Wang(2014)}]{LeeWang}
Lee JY, Wang B (2014) Future change of global monsoon in the cmip5. Climate
  Dynamics 42:101 -- 119

\bibitem[{Li et~al(2010)Li, Sheffield, and Wood}]{Li}
Li H, Sheffield J, Wood EF (2010) Bias correction of monthly precipitation and
  temperature fields from {I}ntergovernmental {P}anel on {C}limate {C}hange
  {AR}4 models using equidistant quantile matching. J Geophys Res 115:D10,101,
  \doi{10.1029/2009JD012882}

\bibitem[{Liepert and Lo(2013)}]{Liepert2}
Liepert BG, Lo F (2013) {CMIP}5 updates of ``inter-model variability and biases
  of the global water cycle in {CMIP}3 coupled climete models''. Environmental
  Research Letters 8, \doi{10.1088/1748-9326/8/2/029401}

\bibitem[{Lin(2007)}]{Lin}
Lin JL (2007) The double-itcz problem in {IPCC AR}4 coupled {GCM}s:
  OceanÐatmosphere feedback analysis. J Climate 20:4497 -- 4525

\bibitem[{Lucarini et~al(2008)Lucarini, Danihlik, Kriegerova, and
  Speranza}]{Danube}
Lucarini V, Danihlik R, Kriegerova I, Speranza A (2008) Hydrological cycle in
  the {D}anube basin in present-day and {XXII} century simulations by {IPCCAR}4
  global climate models. J Geophys Res 113:D09,107, \doi{10.1029/2007JD009167}

\bibitem[{Meehl et~al(2007)Meehl, Stocker, Collins, Friedlingstein, Gaye,
  Gregory, Kitoh, Knutti, Murphy, Noda, Raper, Watterson, Weaver, and
  Zhao}]{Meehl}
Meehl G, Stocker T, Collins W, Friedlingstein P, Gaye AT, Gregory JM, Kitoh A,
  Knutti R, Murphy JM, Noda A, Raper SCB, Watterson IG, Weaver AJ, Zhao ZC
  (2007) {IPCC} {C}limate {C}hange 2007: The physical science basis. eds. {S}.
  {S}olomon et al., Cambridge Univ. Press, pp 747--846

\bibitem[{Mehran et~al(2014)Mehran, AghaKouchak, and Phillips}]{Mehran}
Mehran A, AghaKouchak A, Phillips TJ (2014) Evaluation of {CMIP}5 continental
  precipitation simulations relative to satellite-based gauge-adjusted
  observations. Journal of Geophysical Research 119:1695 -- 1707,
  \doi{10.1002/2013JD021152}

\bibitem[{Mitchell and Jones(2005)}]{Mitchell}
Mitchell TD, Jones PD (2005) An improved method of constructing a database of
  monthly climate observations and associated high-resolution grids. Int J
  Climatol 25:693--712

\bibitem[{Noake et~al(2012)Noake, Polson, Hegerl, and Zhang}]{Noake}
Noake K, Polson D, Hegerl G, Zhang X (2012) Changes in seasonal land
  precipitation during the latter twentieth-century. Geophysical Research
  Letters 39:L03,706, \doi{10.1029/2011GL050405}

\bibitem[{Rathmann et~al(2013)Rathmann, Yang, and Kaas}]{Rathmann}
Rathmann NM, Yang S, Kaas E (2013) Tropical cyclones in enhanced resolution
  cmip5 experiments. Climate Dynamics \doi{10.1007/s00382-013-1818-5}

\bibitem[{Rohr et~al(2013)Rohr, Manzoni, Feng, Menezes, and Porporato}]{Rohr}
Rohr T, Manzoni S, Feng X, Menezes RSC, Porporato A (2013) Effect of rainfall
  seasonality on carbon storage in tropical dry ecosystems. J Geophys Res
  Biogeosci 118:1156 -- 1167, \doi{10.1002/jgrg.20091}

\bibitem[{Sarojini et~al(2012)Sarojini, Stott, Black, and Polson}]{Beena}
Sarojini BB, Stott PA, Black E, Polson D (2012) Fingerprints of changes in
  annual and seasonal precipitation from {CMIP}5 models over land and ocean.
  Geophys Res Lett 39:L21,706, \doi{doi:10.1029/2012GL053373}

\bibitem[{Schertzer and Lovejoy(1987)}]{Lovejoy}
Schertzer D, Lovejoy S (1987) Physical modeling and analysis of rain and clouds
  by anisotropic scaling multiplicative processes. J Geophys Res 92(D8):9693 --
  9714

\bibitem[{Schneider et~al(2013)Schneider, Becker, Finger, Meyer-Christoffer,
  Ziese, and Rudolf}]{GPCC}
Schneider U, Becker E, Finger P, Meyer-Christoffer A, Ziese M, Rudolf B (2013)
  {GPCC}'s new land surface precipitation climatology based on
  quality-controlled in situ data and its role in quantifying the global water
  cycle. Theoretical and Applied Climatology \doi{10.1007/s00704-013-0860-x}

\bibitem[{Seager et~al(2010)Seager, Naik, and Vecchi}]{Seager}
Seager R, Naik N, Vecchi G (2010) Thermodynamic and dynamic mechanisms for
  large-scale changes in the hydrological cycle in response to global warming.
  Journal of Climate 23:4651--4668

\bibitem[{Seager et~al(2013)Seager, Ting, Li, Naik, Cook, Nakamura, and
  Liu}]{Seager2}
Seager R, Ting M, Li C, Naik N, Cook B, Nakamura J, Liu H (2013) Projections of
  declining surface-water availability for the southwestern united states.
  Nature Climate Change 3:482--486

\bibitem[{Seth et~al(2013)Seth, Rauscher, Biasutti, Giannini, Camargo, and
  Rojas}]{Seth}
Seth A, Rauscher SA, Biasutti M, Giannini A, Camargo SJ, Rojas M (2013) {CMIP}5
  projected changes in the annual cycle of precipitation in monsoon regions. J
  Climate 26:7328 -- 7351

\bibitem[{Shukla and Paolino(1983)}]{Shukla}
Shukla J, Paolino DA (1983) The {S}outhern {O}scillation and long range
  foresting of the summer monsoon rainfall over {I}ndia. Mon Weath Rev
  111:1830--1837

\bibitem[{Sillmann et~al(2013)Sillmann, Kharin, Zhang, Zwiers, and
  Bronough}]{Sillman}
Sillmann J, Kharin VV, Zhang Z, Zwiers FW, Bronough D (2013) Climate extremes
  indecision the {CMIP}5 multimodel ensembles: {P}art {I}. {M}odel evaluation
  in the present climate. Journal of Geophysical Research 118(4):1716 -- 1733

\bibitem[{Sperber et~al(2013)Sperber, Annamalai, Kang, Kitoh, Moise, Turner,
  Wang, and Zhou}]{Sperber}
Sperber KR, Annamalai H, Kang IS, Kitoh A, Moise A, Turner A, Wang B, Zhou T
  (2013) The asian summer monsoon: an intercomparison of {CMIP}5 vs. {CMIP}3
  simulations of the late 20th century. Climate Dyanmics 41:2711--2744

\bibitem[{Swart and Fyfe(2012)}]{Swart}
Swart NC, Fyfe JC (2012) Observed and simulated changes in the southern
  hemisphere surface westerly wind-stress. Geophys Res Lett 39:L16,711,
  \doi{10.1029/2012GL052810}

\bibitem[{Taylor(2001)}]{TaylorDiag}
Taylor KE (2001) Summarizing multiple aspects of model performance in a single
  diagram. J Geophys Res 106:7183 -- 7192

\bibitem[{Taylor et~al(2012)Taylor, Stouffer, and Meehl}]{Taylor}
Taylor KE, Stouffer RJ, Meehl GA (2012) An overview of {CMIP5} and the
  experiment design. Bull Amer Meteor Soc 93:485--498

\bibitem[{Trenberth et~al(2000)Trenberth, Stepaniak, and Caron}]{Trenberth00}
Trenberth KE, Stepaniak DP, Caron JM (2000) The global monsoon as seen through
  the divergent atmospheric circulations. J Climate 13:3969 -- 3993

\bibitem[{Turner and Annamalai(2012)}]{Turner}
Turner A, Annamalai H (2012) Climate change and the {S}outh {A}sian summer
  monsoon. Nature Climate Change 2:587--595

\bibitem[{Vecchi and Soden(2007)}]{Vecchi}
Vecchi GA, Soden BJ (2007) Global warming and the weakening of tropical
  circulation. J Climate 20:4316 -- 4340

\bibitem[{Vellinga et~al(2013)Vellinga, Arribas, and Graham}]{WAF}
Vellinga M, Arribas A, Graham R (2013) Seasonal forecasts for regional onset of
  the west african monsoon. Climate Dynamics 40(11):3047 -- 3070

\bibitem[{van Vuuren et~al(2011)van Vuuren, Edmonds, Kainuma, Riahi, Thomson,
  Hibbard, Hurtt, Kram, Krey, Lamarque, Masui, Meinshausen, Nakicenovic, Smith,
  and Rose}]{RCP}
van Vuuren DP, Edmonds J, Kainuma M, Riahi K, Thomson A, Hibbard K, Hurtt GC,
  Kram T, Krey V, Lamarque JF, Masui T, Meinshausen M, Nakicenovic N, Smith SJ,
  Rose SK (2011) The representative concentration pathways: an overview.
  Climatic Change 109:5--31

\bibitem[{Walsh and Lawler(1981)}]{Walsh}
Walsh RPD, Lawler DM (1981) Rainfall seasonality. Weather 36:201--208

\bibitem[{Wang and Ding(2008)}]{Ding}
Wang B, Ding Q (2008) Global monsoon: dominant mode of annual variation in the
  tropics. Dynamics of Atmospheres and Oceans 44:165 -- 183

\bibitem[{Wang and Fan(1999)}]{Wang3}
Wang B, Fan Z (1999) Choice of south asian summer monsoon indices. Bull Amer
  Meteor Soc 80:629--638

\bibitem[{Wang et~al(2011)Wang, Kim, Kikuchi, and Kitoh}]{Wang}
Wang B, Kim HJ, Kikuchi K, Kitoh A (2011) Diagnostic metrics for evaluation of
  annual and diurnal cycles. Climate Dynamics 37:941--955

\bibitem[{Webster and Yang(1992)}]{Webster}
Webster PJ, Yang S (1992) {M}onsoon and {ENSO}: {S}electively interactive
  systems. Quart J Roy Meteor Soc 118:877--926

\bibitem[{Xie and Arkin(1997)}]{Xie}
Xie P, Arkin PA (1997) Global precipitation: a 17-year monthly analysis based
  on gauge observations, satellite estimates, and numerical model outputs. Bull
  Amer Meteor Soc 78:2539--2558

\bibitem[{Xie et~al(2003)Xie, Janowiak, Arkin, Adler, Gruber, Ferraro, Huffman,
  and Curtis}]{Xie2}
Xie P, Janowiak JE, Arkin PA, Adler R, Gruber A, Ferraro R, Huffman GJ, Curtis
  S (2003) {GPCP} pentad precipitation analyses: an experimental dataset based
  on gauge observations and satellite estimates. J Climate 16:2197 -- 2214

\bibitem[{Zappa et~al(2013)Zappa, Shaffrey, and Hodges}]{Zappa}
Zappa G, Shaffrey LC, Hodges KI (2013) The ability of cmip5 models to simulate
  north atlantic extratropical cyclones. J Climate 26(15):5379--5396,
  \doi{10.1175/JCLI-D-12-00501.1}

\bibitem[{Zhang et~al(2007)Zhang, Zwiers, Hegerl, Lambert, Gillett, Stott, and
  Nosawa}]{Zhang}
Zhang X, Zwiers FW, Hegerl GC, Lambert FH, Gillett NP, Stott PA, Nosawa T
  (2007) Detection of human influence on twentieth-century precipitation
  trends. Nature 448:461--465

\end{thebibliography}

%%\begin{thebibliography}{}
%%\bibitem{RefJ}
%%Author, Article title, Journal, Volume, page numbers (year)
%%\bibitem{RefB}
%%Author, Book title, page numbers. Publisher, place (year)
% etc
%%\end{thebibliography}

\end{document}